\documentclass[11pt,onecolumn,oneside]{osajnl}

\journal{jocn} 

\title{On Secrecy Performance of Mixed $ \alpha-\eta-\mu $ and M{\'a}laga RF-FSO Variable Gain Relaying Channel}

\author[1]{Nandita Swanan Mandira}
\author[1]{Milton Kumar Kundu}
\author[2]{Sheikh Habibul Islam}
\author[3]{A. S. M. Badrudduza}
\author[4]{Imran Shafique Ansari}

\affil[1]{Department of Electrical \& Computer Engineering, Rajshahi University of Engineering \& Technology (RUET), Rajshahi-6204, Bangladesh}
\affil[2]{ Department of Electrical \& Electronic Engineering, RUET }
\affil[3]{Department of Electronics \& Telecommunication Engineering, RUET}
\affil[4]{James Watt School of Engineering, University of Glasgow, Glasgow G12 8QQ, United Kingdom}

\begin{abstract}
With the completion of standardization of fifth-generation (5G) networks, the researchers have begun visioning sixth-generation (6G) networks that are predicted to be human-centric. Hence, similar to 5G networks, besides high data rate, providing secrecy and privacy will be the center of attention by the wireless research community. To support the visions beyond 5G (B5G) and 6G, in this paper we propose a secure radio frequency (RF)-free space optical (FSO) mixed framework under the attempt of wiretapping by an eavesdropper at the RF hop. We assume the RF links undergo $\alpha-\eta-\mu$ fading whereas the FSO link exhibits a unified M\'alaga turbulence model with pointing error. The secrecy performance is evaluated by deducing expressions for three secrecy metrics i.e. average secrecy capacity, secure outage probability, and probability of non-zero secrecy capacity in terms of univariate and bivariate Meijer's $G$ and Fox's $H$ functions. We further capitalize on these expressions to demonstrate the impacts of fading, atmospheric turbulence, and pointing errors and show a comparison between two detection techniques (i.e. heterodyne detection (HD) and intensity modulation with direct detection (IM/DD)) that clearly reveals better secrecy can be achieved with HD technique relative to the IM/DD method. The inclusion of generalized fading models at the RF and FSO hops offers unification of several classical scenarios as special cases thereby exhibiting a more generic nature relative to the existing literature. Finally, all the analytical results are corroborated via Monte-Carlo simulations. 

\quad

\end{abstract}

\begin{document}

\maketitle

\section{Introduction}
\subsection{Background}
The ever increasing demand for higher data rates at lower costs has been the motivation for researchers to explore new data transfer technologies. As a result, the data transfer speed is increasing with every generation of the communication systems. In this era of fifth-generation (5G) communications and beyond (B5G), technicians are targeting to achieve a data rate of 1 Gbps \cite{andrews2014will}. To achieve this very high-speed data transfer, free-space optical (FSO) technologies can be a great medium of data communication.

As 5G is already deployed in many areas of the world, researchers have already begun their work on 6G communications where the security of information and privacy of the users is going to be prioritized. The authors in \cite{dang2020should} suggested that as the encryption method is failing to provide the necessary security against the powerful computing systems, physical layer security (PLS) can be a better option for ensuring secrecy. They also suggested that visible light communications (in other words FSO communications) can be put to use to achieve this goal. The application of FSO via UAVs can also be a great solution to the bandwidth requirements and security problem that arises in 6G communications \cite{seo2020combined}. At present, the FSO network is being applied to military communications, metropolitan area network (MAN) extensions, and to resolve the “last mile access” problem \cite{islam2021impact}. Although the FSO networking system has numerous advantages, it is yet to be considered for deployment at a large scale because of its short-range transmission capacity. This disadvantage can be easily removed by the hybridization of radio frequency (RF) and FSO networks that will provide us all the advantages provided by both RF and FSO systems. Such hybrid RF-FSO communication systems can transmit data at long distances with high data rates \cite{sharma2019effect}.

\subsection{Literature Survey}
The advantages presented by RF-FSO networks has caught the attention of researchers that has led them to analyze the possibility of real-life applications of such hybrid systems. The performance of RF-FSO networks for the simplest of combinations of communication nodes (assuming only a single source, an amplify-and-forward (AF) relay, and a receiver) was analyzed in \cite{1anees2015performance, ansari2013impact, upadhya2018relay, 7881143} considering the effects of turbulence and pointing errors at the FSO link. A similar model was also analyzed in \cite{anees2015performance, wang2018performance} but in these works, the authors considered decode-and-forward (DF) relaying scheme instead of the AF relay. A cognitive radio network was also employed with an underlay RF-FSO system in \cite{6950766,6952039,7883900,erdogan2019performance} as it is a viable solution that can be used in metropolitan networking. The authors in \cite{alimi2017analysis} designed a RF-FSO cloud computing-based radio access network (CC-RAN) for 5G communication with Rayleigh and Gamma-Gamma (GG) fading models at the RF and FSO hop, respectively, considering multiple senders. As generalized fading models allow us the versatility of using multiple multipath fading models without any complex mathematical manipulation, researchers started employing them in RF-FSO networks for enhancing its performance. Due to this reason, another CC-RAN was evaluated with $\kappa-\mu$ shadowed fading and exponentiated Weibull (EW) fading models at the RF and FSO hops, respectively, considering a single \cite{yi2019performance} and multiple users \cite{yi2020performance}.

Researchers have also considered some more complex communication scenarios within RF-FSO communication technologies to reduce the disadvantages caused by pointing error or turbulence at FSO hop. Authors in \cite{zedini2014performance, zhao2017performance, tonk2020mixed} evaluated a scenario where multiple sources are sending data to the relay and the relay is multiplexing and retransmitting that information to the receiver through the FSO link. An analogous model was considered in \cite{feng2016performance} where the information from multiple sources is re-transmitted via a relay to the receiver using time division multiple access (TDMA) techniques. The advantages of transmit antenna selection (TAS) scheme was perused in \cite{odeyemi2019performance} over a RF-FSO system considering multiple receivers. A data transfer scenario with both direct and relay assisted links was explored in \cite{wang2019performance, asgari2019performance} where the relay assisted link was a RF-FSO link with both sender and receiver having multiple antennas and multiple receiver apertures, respectively \cite{wang2019performance}. The capacity of a FSO link can be enhanced by employing a multiple-input multiple-output (MIMO) configuration that was shown in \cite{han2018performance, chen2017multi}. In \cite{torabi2019performance}, authors described a very unique scenario in a RF-FSO network where both the relay and the destination had RF antennas and optical apertures so the relay was able to send information using RF or optical signals and the receiver could receive both the signals. A comparison was then made between single-input single-output (SISO) and dual-input dual-output (DIDO) based relay systems applied to a RF-FSO configuration where the authors concluded that DIDO outperforms SISO significantly \cite{zhang2018performance}. The relay selection scheme was employed in \cite{erdogan2019joint} to enhance the performance of a system having multiple sources. A few works also show a reverse technology where the authors have considered FSO technology as the source to relay link and RF technology for relay to destination link \cite{odeyemi2019selection, jing2017performance, 1amirabadi2019performance, amirabadi2019performance} where the authors in \cite{jing2017performance, 1amirabadi2019performance, amirabadi2019performance} have considered multiple receivers at the RF hop.

In the present era, ensuring high data speed is not enough if the transmission of information is insecure. In the early stages of wireless communications, the cryptographic approach was more popular among the researchers. But due to the requirement of high power and complexity, this method is losing its popularity \cite{van2018physical}. On the other hand, PLS has emerged as an auspicious paradigm for information security against malicious attackers. Researchers around the world have started analyzing the PLS for RF-FSO communication technology. A general eavesdropping scenario for RF-FSO systems was assumed in \cite{lei2017secrecy, yang2018physical, sarker2020secrecy, islam2020secrecy} where the eavesdropper is trying to overhear the legitimate information through the RF link. Similar RF-FSO models have also been subjected to the secrecy analysis considering transmit antenna selection (TAS) schemes \cite{lei2020secure} and maximal ratio combining (MRC) technique at both relay and eavesdropper \cite{lei2018secrecy}. An opportunistic user scheduling (OUS) mechanism in RF-FSO network with multi-antenna relay and eavesdropper was interpreted in \cite{abd2016security}. Authors in \cite{abd2017effect} made a trade-off between security and reliability for an OUS relay network with power allocation. Security of the system can be hampered if multiple eavesdroppers are present in the system that was shown in \cite{odeyemi2018physical}. Researchers have found that this problem can be solved by using the partial relay selection scheme \cite{odeyemi2019security}.

\subsection{Motivation and Contributions}
In physical layer communication over both RF and FSO links, a variation takes place to the transmitted signal due to the random physical properties of the channels through which the information is transmitted. Since this randomness of the wireless channels can not be predicted beforehand, and the channels between the transmit and receive nodes are changing at each and every instant, hence assuming a generalized scenario at these links has proven to be more practical. This is why the researchers around the world started considering generalized channels for evaluating the performance of wireless communication systems. In a mixed RF-FSO system, although some generalized RF channels, e.g. $\eta-\mu$, hyper Gamma, and generalized Gamma \cite{yang2018physical, sarker2020secrecy, islam2020secrecy, 6692669, 7145973, 10754/134733}, etc. have been considered, these channels do not cover all the physical aspects of communication through a wireless medium. To remedy this situation, the generalized $\alpha-\eta-\mu$ channel is considered in this work as the RF link that incorporates the non-homogeneity and the non-linearity nature of the fading channels with no line-of-sight (LOS) component \cite{moualeu2018physical}. Furthermore, the $\alpha-\eta-\mu$ fading distribution includes $\alpha-\mu$ and $\eta-\mu$  generalized fading scenarios along with their special multipath cases such as Nakagami-$m$, Nakagami-$q$, Rayleigh, and Weibull \cite{badarneh2015error}. On the other hand, the FSO link of this work is considered to follow the well-known unified M{\'a}laga model with pointing error impairments that also represents a generalized scenario by including GG, Log-Normal, Rice-Nakagami, etc. distributions as its special cases \cite{6966082, 7145711}. As a result, the proposed ($\alpha-\eta-\mu$)-M{\'a}laga RF-FSO mixed distribution can cover a wide variety of mixed configurations that can handle any practical scenario more precisely relative to the existing literature.

\begin{itemize}
\item First of all, we derive the expressions of dual-hop cumulative distribution functions (CDFs) for the proposed RF-FSO system utilizing the probability density functions (PDFs) of $\alpha-\eta-\mu$ and M{\'a}laga fading distributions considering variable gain relaying technique. The derived CDFs are completely unique and can be utilized to derive the dual-hop CDFs of \cite{yang2018physical} and \cite{islam2020secrecy} as our special cases.

\item We analyze the secrecy performance of the proposed system incorporating the impact of atmospheric turbulence, fading parameters, and pointing error. The expressions of average secrecy capacity (ASC), lower bound of secrecy outage probability (SOP), and probability of non-zero secrecy capacity (PNSC) are derived in terms of well-known univariate and bivariate Meijer's $G$ and Fox's $H$ functions. The derived expressions are novel and can be utilized to replicate the results of \cite{lei2017secrecy, yang2018physical}. A comparative study is also presented between HD and IM/DD techniques for better clarification.

\item Monte-Carlo (MC) simulations are provided for verifying the accuracy of the derived analytical expressions of ASC, lower bound of SOP, and PNSC.

\item Our analysis suggests the system security boosts up with the reduction of severity of atmospheric turbulence, enhancement of the values of fading parameters, and increment in the value of pointing error. Moreover, the PLS of the considered model always exhibits superior performance under the HD technique relative to the IM/DD technique.
\end{itemize}

\subsection{Organization}

The rest of this paper is organized in the following manner. Section II narrates the proposed system model along with mathematical formulation. Section III derives the closed-form expressions for performance parameters (ASC, lower bound of SOP, and PNSC). Section IV demonstrates numerical results for these derived performance parameters via utilizing graphical representations followed by summarizing the work in Section VI.

\section{System Model and Problem Formulation}
\label{system}

\begin{figure}[!ht]
\vspace{-15mm}
    \centerline{\includegraphics[width=0.7\textwidth]{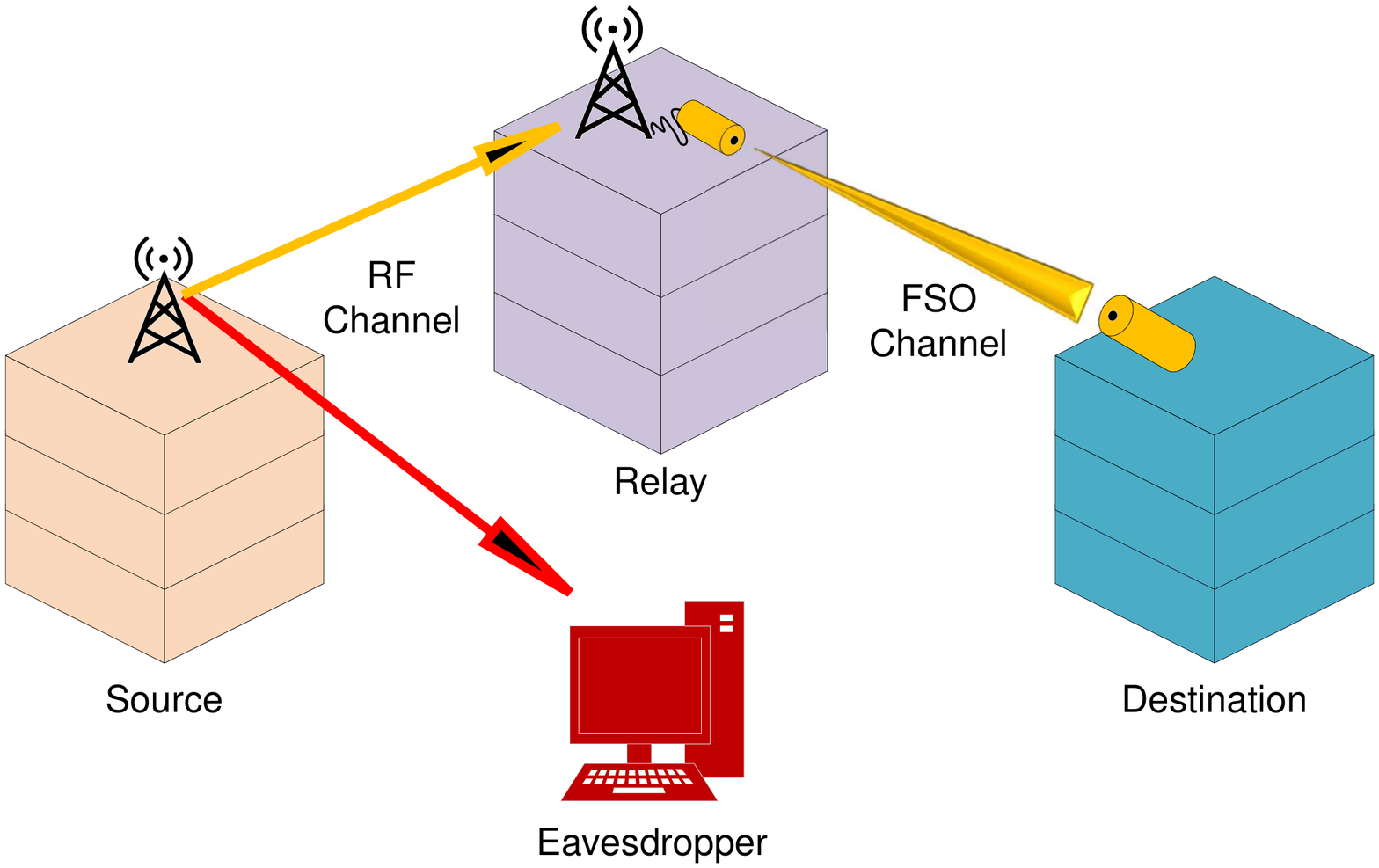}}
    \vspace{-15mm}
    \caption{Dual hop RF-FSO system including a source ($A_s$), a relay ($B_r$), a destination ($C_d$), and an eavesdropper ($E_v$).}
   \label{fig.1}
\end{figure}
Our proposed system model is demonstrated in Fig. \ref{fig.1}, which includes a source ($A_s$), a relay ($B_r$), a destination ($C_d$), and an eavesdropper ($E_v$). Each of $A_s$ and $E_{v}$ are equipped with a single antenna whereas $B_{r}$ and $C_d$ have a single aperture for transmission and reception of optical signals, respectively. We consider the destination receives transmitted RF signal from source in optical form whereby the RF to optical conversion is performed at the relay node. On the other hand, eavesdropper takes advantage of the time varying nature of RF link while it is on a continuous hunt to steal information via source to eavesdropper ($A_s-E_v$) link. The source to relay ($A_s-B_r$) and $A_s-E_v$ links are assumed to follow independently distributed $\alpha-\eta-\mu$ fading whereas the relay to destination ($B_r-C_d$) link experiences M{\'a}laga turbulence impaired with pointing error. The motivation behind selecting these two distributions is that both $\alpha-\eta-\mu$ and M{\'a}laga turbulence fading models provide excellent match with the experimental data \cite{salameh2019end, arya2020multiuser}. Moreover, the $\alpha-\eta-\mu$ fading model has the highest accuracy to represent the variation in transmitted signals over various different multipath fading channels \cite{salameh2019end}.

\subsection{RF Channel Model }
Denoting the channel gain of $A_s-B_r$ link as $a_{sr}$ and the corresponding SNR as $\gamma_{r}=\frac{P_s}{N_r}\|a_{sr}^{2}\|$, where $P_s$ and $N_r$ are the transmit power of $A_s$ and noise power at $B_r$, respectively. We can express the PDF of $\gamma_{r}$ as \cite{moualeu2018physical}
\begin{align}
    f_{r}(\gamma)&= \frac{S_r}{{\omega}_r^{\beta_r}}\gamma^{\beta_r-1}e^{-A_r(\frac{\gamma}{{\omega}_r})^{\tilde{\alpha}_r}} I_{\mu_r-\frac{1}{2}}\biggl(P_r\left(\frac{\gamma}{{\omega}_r}\right)^{\tilde{\alpha_r}}\biggl),
\end{align}
where $S_r =\frac{\sqrt{\pi}{\tilde{\alpha}_r}\mu_r^{\mu_r+\frac{1}{2}}(\eta_r-1)^{\frac{1}{2}-\mu_r}(\eta_r+1)^{\mu_r+\frac{1}{2}}}{\eta_r\Gamma(\mu_r)}$, $A_r=\frac{\mu_r(1+\eta_r)^{2}}{2\eta_r}$, $\beta_r={\tilde{\alpha}_r}({\mu_r+\frac{1}{2})}$, $P_r=\frac{\mu_r({\eta_r}^{2}-1)}{2\eta_r}$, and $\tilde{\alpha}_r=\frac{\alpha_r}{2}$. Here, the average SNR is denoted by ${\omega}_r$, $I_r(.)$ represents the first-order modified Bessel function as defined in \cite[Eq.~(9.6.20),]{abramowitz1972handbook}, non-linearity parameter is denoted by ${\alpha}_r$, the number of multipath clusters is denoted by $\mu_r$, and the ratio between in-phase and quadrature scattered wave components is denoted by $\eta_r$. Making use of \cite[Eq.~(8.445),]{gradshteyn2014table} and employing mathematical simplifications and manipulations, we can write $f_{r}(\gamma)$ as
\begin{align}
\label{a2}
f_{r}(\gamma)=\sum_{N_1=0}^{\infty}u_{1}\gamma^{u_{3}}e^{{-u_{2}}{\gamma^{\tilde{\alpha}_r}}},
\end{align}
where $u_{1}$=$\frac{ S_r \left(\frac{P_r}{{2\omega}_r^{{\tilde{\alpha}_r}}}\right)^{{\mu_r-\frac{1}{2}}+2N_1}}{N_1!{\omega}_r^{\beta_r}\Gamma({\mu_r+\frac{1}{2}}+N_1)}$, $u_{2}$=$(\frac{A_r}{{\omega}_r^{\tilde{\alpha}_r}})$,  and $u_{3}$=${\tilde{\alpha}_r}({\mu_r-\frac{1}{2}}+2N_1)+\beta_r-1$. We define the CDF of $\gamma_r$ as
\begin{align}
\label{a3}
F_{r}(\gamma)=\int_{0}^{\gamma_r}f_{r}(\gamma)d\gamma.
\end{align}
Placing \eqref{a2} into \eqref{a3} and utilizing \cite[Eqs.~(3.381.8) and (8.352.6,]{gradshteyn2014table}, we obtain the final expression for CDF of $\gamma_r$ as
\begin{align}
\label{a4}
F_{r}(\gamma)=1-\sum_{N_1=0}^{\infty}\sum_{t_1=0  }^{W_1-1}u_{4}{e^{-u_{2}\gamma^{\tilde{\alpha}_r}}\gamma^{\tilde{\alpha}_rt_1}},
\end{align}
where $W_1=\frac{u_{3}+1}{\tilde{\alpha}_r}$ and $u_{4}=(W_1-1)!\frac{u_{1}}{\tilde{\alpha}_r{u_{2}}^{W_1}}\frac{{u_{2}}^{t_1}}{t_1!}$.

\subsection{FSO Channel Model }
Denoting the channel gain of $B_r-C_d$ link as $a_{rd}$, we can define the corresponding SNR as $\gamma_d=\frac{P_{r}}{N_{d}}\|a_{rd}^{2}\|$, where $P_{r}$ and $N_d$ symbolize the transmit power from relay and noise power at the destination, respectively.
We assume $\gamma_{d}$ follows the unified M\'alaga distribution with pointing error, the PDF of which can be expressed as \cite{ansari2015performance}
\begin{align}
\label{a5}
   f_d (\gamma)=\frac{{\varepsilon}^2 A_d}{2^r \gamma}\sum _{\tilde {m}_d=1}^{{\beta}_d} b_d G_{1,3}^{3,0}\left[B_d \left(\frac{\gamma}{u_r}\right)^{\frac{1}{r}}\biggl|
\begin{array}{c}
 {\varepsilon} ^2+1 \\
 {\varepsilon} ^2,{{\alpha}_d} ,{\tilde {m}_d} \\
\end{array}
\right],
\end{align}
where
\begin{align}
    \nonumber
    &A_d=\frac{2{\alpha_d}^{\frac{\alpha_d}{2}}}{g_d^{1+\frac{\alpha_d}{2}}\Gamma(\alpha_d)}\biggl(\frac{g_d\beta_d}{g_d\beta_d+\Omega_d}\biggl)^{\beta_d+\frac{\alpha_d}{2}},
    \\
    \nonumber
    &B_d=\frac{\varepsilon^2\alpha_d\beta_d(g_d+\Omega_d)}{(\varepsilon^2+1)(g_d\beta_d+\Omega_d)},
    \\
    \nonumber
    &a_d=\binom{\beta_d-1}{\tilde {m}_d-1}\frac{(g_d\beta_d+\Omega_d)^{1-\frac{\tilde {m}_d}{2}}}{(\tilde {m}_d-1)!}{\biggl(\frac{\Omega_d}{g_d}\biggl)^{\tilde {m}_d-1}}\biggl(\frac{\alpha_d}{\beta_d}\biggl)^{\frac{\tilde {m}_d}{2}}, \text{ and}
    \\
    \nonumber
    &b_d=\alpha_d\biggl(\frac{\alpha_d\beta_d}{g_d\beta_d+\Omega_d}\biggl)^{-\frac{\alpha_d+\tilde {m}_d}{2}}.
\end{align}
In \eqref{a5}, the electrical SNR is denoted by $u_r$, where the optical signal detection system is referred by $r$ i.e. $r=1$ is meant for HD technique with $u_{HD}=\bar{\gamma}_{d}$ and $r=2$ is meant for IM/DD technique with $u_{IM/DD}=\frac{\alpha_d{\varepsilon}^2(g_d+\Omega_d)({\varepsilon}^2+2)}{(\varepsilon^2+1)^2(\alpha_d+1)[2g_d(g_d+2\Omega_d)+{\Omega_d}^2(1+\frac{1}{\beta_d})]}\bar{\gamma}_{d}$ \cite{ansari2015performance}, and $\overline{\gamma}_{d}$ is the average SNR. The effective number of large scale cells during scattering process is denoted by $\alpha_d$, natural number $\beta_d$ represents fading parameter's amount, the average power of the scattering component in the off-axis eddies is denoted by $g_d=\mathbb{E}[{|U_{s}^g|}^2]=2b_d(1-\rho)$, $2b_d=\mathbb{E}[{{|U_{s}^c|}^2}+{{|U_{s}^g|}^2}]$ denotes the average power of scattering components, the average power from the coherent contributions is represented by $\Omega_d$, where $\Omega_d=\Omega+2b_d\rho+2\sqrt{2b_d\rho\Omega}\cos(\Theta_A-\Theta_B)$ refers to the average power regarding coherent advantages, regular power through LOS component is defined by $\Omega=\mathbb{E}[{|U_{los}|}^2]$, the deterministic phases of LOS and coupled-to-LOS spread terms are referred by $\Theta_A$ and $\Theta_B$, respectively. Here, the amount of coupling between scattering components and LOS component is described with a range of $0\leq \rho\leq1$, the quantity of scattering power coupled LOS component is denoted by $0\leq \rho\leq1$ \cite{ansari2015performance}, and the ratio of the equivalent beam radius to the pointing error displacement standard deviation (jitter) at the receiver is referred by $\varepsilon$. The Meijer's $G$ function is symbolized by $G[.]$ that is defined in \cite{gradshteyn2014table}. The CDF of $\gamma_{d}$ is given by \cite[Eq.~(9),]{ansari2015performance}
\begin{align}
\label{a6}
   F_d (\gamma )=\sigma \sum _{\tilde {m}_d=1}^{\beta_d } c_d G_{r+1,3r+1}^{3r,1}\left[\frac{F }{u _r}\gamma \biggl|
\begin{array}{c}
 1,l_1 \\
 l_2,0 \\
\end{array}
\right],
\end{align}
where $\sigma$=$\frac{\varepsilon^2 A_d}{2^r (2\pi)^{r-1}}$, $F$=$\frac{{B_d}^r}{r^{2r}}$, $c_d$=${b_d}r^{\alpha_d+\tilde{m_d}-1}$, $l_1=[\Delta(r,\varepsilon ^2+1)]$ that includes $r$ terms, and $l_2=[\Delta(r,\varepsilon ^2), \Delta(r, \alpha_d),$ $\Delta(r, \tilde {m}_d)]$ that includes $3r$ terms. Here, $\Delta$ defines the series such as $\Delta(a, b)=(\frac{b}{a}, \frac{b+1}{a},......,\frac{b+a-1}{a})$.


\subsection{RF-FSO Mixed Channel Model }

Considering variable gain AF relaying scheme, the equivalent SNR at the destination terminal for the dual-hop RF-FSO mixed system is denoted as $\gamma_o=\frac{\gamma_r \gamma_d}{\gamma_r+\gamma_d+1}\approx \min(\gamma_r, \gamma_d)$ \cite[Eq.~7,]{lei2017secrecy} and the corresponding CDF is given by \cite[Eq.~10,]{saber2018secrecy}
\begin{align}
\nonumber
\label{a9}
   F_{d}(\gamma)&= \Pr \left\{\min  \left\{\gamma _r,\gamma _d\right\}<\gamma  \right\}
   \\
&=F_r(\gamma ) + F_d(\gamma ) - F_r(\gamma )F_d(\gamma ).
\end{align}
By substituting \eqref{a2} and \eqref{a3} into \eqref{a9}, we obtain
\begin{align}
    \label{a11}
     F_{o}(\gamma)&=1-\sum_{N_1=0}^{\infty}\sum_{t_1=0  }^{W_1-1}u_{4}{e^{-u_{2}\gamma^{\tilde{\alpha}_r}}\gamma^{\tilde{\alpha}_rt_1}}
     \left(1- \sigma \sum _{\tilde {m}_d=1}^{\beta_d } c_d G_{r+1,3r+1}^{3r,1}\left[\frac{F }{u_r}\gamma \biggl| 
     \begin{array}{c}
          1,l_1 \\
          l_2,0 \\
     \end{array}
     \right]\right).
\end{align}


\subsection{RF Channel Model}

We consider the eavesdropper channel also undergoes $\alpha-\eta-\mu$ fading. Hence, denoting the channel gain of $A_s-E_v$ link as $a_{sv}$ and corresponding SNR as $\gamma_{v}=\frac{P_s}{N_v}\|a_{sv}^{2}\|$, where $N_v$ is the noise power at $E_v$, we can express the PDF of $\gamma_{v}$ as \cite{moualeu2018physical}
\begin{align}
\label{aa}
    f_{v}(\gamma)&= \frac{S_v}{{\omega}_v^{\beta_v}}\gamma^{\beta_v-1}e^{-A_v(\frac{\gamma}{{\omega}_v})^{\tilde{\alpha}_v}}I_{\mu_v-\frac{1}{2}}\biggl(P_v(\frac{\gamma}{{\omega}_v})^{\tilde{\alpha_v}}\biggl),
\end{align}
where $S_v =\frac{\sqrt{\pi}{\tilde{\alpha}_v}\mu_v^{\mu_v+\frac{1}{2}}(\eta_v-1)^{\frac{1}{2}-\mu_v}(\eta_v+1)^{\mu_v+\frac{1}{2}}}{\eta_v\Gamma(\mu_v)} $, $A_v=\frac{\mu_v(1+\eta_v)^{2}}{2\eta_v}$, $\beta_v={\tilde{\alpha}_v}({\mu_v+\frac{1}{2})}$, $P_v=\frac{\mu_v({\eta_v}^{2}-1)}{2\eta_v}$, $\tilde{\alpha}_v=\frac{\alpha_v}{2}$,  ${\omega}_v$ denotes the average SNR of eavesdropper channel, $\alpha_v$ denotes the non-linearity parameter, the number of multipath clusters are represented by $\mu_v$, and $\eta_v$ denotes the ratio between in phase and quadrature scattered wave components. Utilizing \cite[Eq.~(8.445),]{gradshteyn2014table} and performing some algebraic manipulations, we simplify the PDF in \eqref{aa} as
\begin{align}
\label{a12}
f_{v}(\gamma)=\sum_{N_2=0}^{\infty}q_1\gamma^{q_3}e^{{-q_2}{\gamma^{\tilde{\alpha}_v}}},
\end{align}
where $q_1$=$\frac{{S_v}(\frac{P_v}{{2{\omega}_v}^{\tilde{\alpha}_v}})^{{\mu_v-\frac{1}{2}}+2N_2}}{N_2!{\omega}_v^{\beta_v}\Gamma({\mu_v+\frac{1}{2}}+N_2)}$, $q_2$=$(\frac{A_v}{{\omega}_v^{\tilde{\alpha}_v}})$, $q_3$=$\tilde{\alpha}_v({\mu_v-\frac{1}{2}}+2N_2)+{\beta_v-1}$.
Applying \eqref{a3} via integrating \eqref{a12} by utilizing \cite[Eqs.~(3.381.8), and (8.352.6),]{gradshteyn2014table}, we obtain $F_{v}(\gamma)$ as
\begin{align}
\label{a15}
F_{v}(\gamma)=1-\sum_{N_2=0}^{\infty}\sum_{t_2=0  }^{W_2-1}q_{4}{e^{-q_2\gamma^{\tilde{\alpha}_v}}\gamma^{\tilde{\alpha}_vt_2}},
\end{align}
where $W_2$=$\frac{q_3+1}{\tilde{\alpha}_v}$ and $q_4$=$(W_2-1)!\frac{q_1}{\tilde{\alpha}_v{q_2}^{W_2}}\frac{{q_2}^{t_2}}{t_2!}$.


\section{performance metrics}
\label{metric}

In the following subsections, we derive the novel analytical expressions of three secrecy measures i.e. ASC, SOP, and PNSC.

\subsection{Analysis of Average Secrecy Capacity(ASC)}

ASC is one of the most essential secrecy parameters that denotes average of instantaneous secrecy capacity and can be defined as \cite[Eq.~(16),]{lei2017secrecy}
\begin{align}
    \label{a16}
    ASC=\int_0^{\infty } \frac{ 1}{1+\gamma}F_v (\gamma) [1-F_o(\gamma) ]  d\gamma.
\end{align}
Placing \eqref{a11} and \eqref{a15} into \eqref{a16}, the expression of ASC is derived as
    \begin{align}
    \label{a17}
    ASC=\sum_{N_1=0}^{\infty} \sum_{t_1=0  }^{W_1-1}u_{4}\biggl[(\mathcal{S}_1- \sigma \sum _{\tilde {m}_d=1}^{\beta_d } c_d \mathcal{S}_3)-\sum_{N_2=0}^{\infty}\sum_{t_2=0  }^{W_2-1}q_{4}(\mathcal{S}_2-\sigma \sum _{\tilde {m}_d=1}^{\beta_d } c_d \mathcal{S}_4)\biggl],
    \end{align}
where $\mathcal{S}_1$, $\mathcal{S}_2$, $\mathcal{S}_3$, and $\mathcal{S}_4$ are defined as
\begin{subequations}
\begin{align}
\label{19a}
\mathcal{S}_1=&\int_0^{\infty}\frac{e^{-u_{2}\gamma^{\tilde{\alpha}_r}}\gamma^{\tilde{\alpha_r}t_1}}{1+\gamma} d\gamma,
\\
\label{19b}
\mathcal{S}_2=&\int_0^{\infty}\frac{e^{-(u_{2}\gamma^{\tilde{\alpha}_r}+q_2\gamma^{\tilde{\alpha}_v})}\gamma^{\tilde{\alpha_r}t_1+\tilde{\alpha_v}t_2}}{1+\gamma}d\gamma,
\\
\label{19c}
\mathcal{S}_3=& \int_0^{\infty}\frac{e^{-u_{2}\gamma^{\tilde{\alpha}_r}}\gamma^{\tilde{\alpha_r}t_1}}{1+\gamma}
G_{r+1,3r+1}^{3r,1}\left[\frac{F }{u _r}\gamma \biggl|
\begin{array}{c}
1,l_1 \\
l_2,0 \\
\end{array} 
\right] d\gamma,
\\
\label{19d}
\mathcal{S}_4=& \int_0^{\infty}\frac{e^{-(u_{2}\gamma^{\tilde{\alpha}_r}+q_2\gamma^{\tilde{\alpha}_v})}\gamma^{\tilde{\alpha_r}t_1+\tilde{\alpha_v}t_2}}{1+\gamma}
G_{r+1,3r+1}^{3r,1}\left[\frac{F }{u _r}\gamma \biggl|
\begin{array}{c}
1,l_1 \\
l_2,0 \\
\end{array} 
\right] d\gamma,
\end{align}
\end{subequations}
and further derived in Appendix \ref{ASC}.

\subsection{Analysis of Secrecy Outage Probability (SOP)}

SOP signifies the probability that the instantaneous secrecy capacity, $C_s$ \cite{wyner1975wire}, drops below $R_s$, where $R_s$ is the target secrecy rate. Mathematically, we can define SOP as \cite[Eq.~(14),]{lei2018secrecy}
\begin{align}
\nonumber
P_{out}(R_s)&=P_r \{ C_s< R_s\}
\\\label{sop}
 &=\int_0^{\infty} F_o(\theta \gamma +\theta-1)f_v(\gamma) d\gamma,
\end{align}
where $\theta=2^{R_{s}}$ and $R_{s}$ is positive. This definition signifies that a perfect secrecy is achievable if $C_s > R_s$ otherwise the destination will fail to decode the transmitted messages correctly. Due to the mathematical difficulties, deriving exact expression of SOP is challenging. Hence, we adopt an approximation method to derive the lower bound of SOP with variable gain relaying that is given by \cite[Eq.~(7),]{lei2020secure}
\begin{align}
\nonumber
P_{out}(R_s)\geq P_{out}^L(R_s)&=P_r\{ \gamma_o \leq \theta \gamma\}
\\
\label{a27}
 &=\int_0^\infty F_d(\theta \gamma)f_v(\gamma)d\gamma.
\end{align}
Substituting \eqref{a11} as well as \eqref{a12} into \eqref{a27}, SOP can be written as
    \begin{align}
    \label{SOP_final}
    &P_{out}^L(R_s)= 1-\sum_{N_1=0}^{\infty} \sum_{N_2=0}^{\infty}\sum_{t_1=0  }^{W_1-1}u_{4} q_2 \theta^{\tilde{\alpha}t_1}(\mathcal{H}_{1}-\sigma \sum _{\tilde {m}_d=1}^{\beta_d } c_d\mathcal{H}_{2}),
    \end{align}
where $\mathcal{H}_{1}$ and $\mathcal{H}_{2}$ are defined as
\begin{subequations}
\begin{align}
    \mathcal{H}_{1}&=\int_0^{\infty}{\gamma}^{q_3+\tilde{\alpha}t_1}e^{-\kappa\gamma^{\tilde{\alpha}}}d\gamma,
    \\
    \mathcal{H}_{2}&=\int_0^{\infty} {\gamma}^{q_3+\tilde{\alpha}t_1} e^{-\kappa\gamma^{\tilde{\alpha}}} G_{r+1,3r+1}^{3r,1}\left[\frac{F }{u _r}(\theta\gamma) \biggl|
    \begin{array}{c}
     1,l_1 \\
     l_2,0 \\
    \end{array}
    \right] d\gamma,
\end{align}
\end{subequations}
and further derived in Appendix \ref{SOP}.

\subsection{Analysis of Probability of Non-zero Secrecy Capacity (PNSC)}

For ensuring secured transmission of data, $C_s$ must be a positive quantity otherwise the secrecy performance of wireless communication system will cripple. The probability of achieving a positive $C_s$ is defined as \cite{islam2021impact}
\begin{align}
\nonumber
PNSC&=P_r(C_s > 0)
\\\label{a33}
 &=\int_0^{\infty}F_v(\gamma)f_d(\gamma)d\gamma.
\end{align}
By substituting \eqref{a5} and \eqref{a15} into \eqref{a33}, and then on integrating the same, we can derive the closed-form expression of PNSC directly. But this lengthy derivation does not exhibit any useful insight since PNSC can be directly obtained from SOP expression as
\begin{align}
\label{pnsc}
    \text{PNSC}=1- P_{out}(R_s)\big|_{R_{s}=0}.
\end{align}
We utilize \eqref{pnsc} for obtaining the numerical results by demonstrating the impacts of the system parameters on the PNSC performance.

\subsection{Novelty of the Expressions of the Secrecy Metrics}
Although dealing with complicated fading models is a challenging task, this work deals with two generalized fading models at both the RF and FSO links from which a wide number of classical scenarios can be obtained as special cases. Note that composition of ($\alpha-\eta-\mu$)-M\'alaga mixed RF-FSO system is a novel approach since this combination of two in-homogeneous wireless medium is absent in the existing literature. Our derived expressions in \eqref{SOP_final} can showcase the special case of Rayleigh-GG scenario given in \cite[Eqs.~(15),]{el2017physical} with $\alpha_r=\alpha_v=2$, $\mu_r=\mu_v=0.5$, $\eta\rightarrow1$, $\rho=1$, $g_d=0$, and $\Omega_d=1$. Likewise, the (Nakagami-$m$)-GG fading system can be shown as a special case of our system considering $\alpha_r=\alpha_v=2$, $\mu_r=\mu_v=0.5$ when $\eta\rightarrow1$ or $\mu_r=\mu_v=m$ when $\eta\rightarrow0$, $\rho=1$, $g_d=0$, and $\Omega_d=1$ as the expressions presented in \eqref{a17} and \eqref{SOP_final} agree with the results of \cite[Eqs.~(13) \& (20),]{lei2017secrecy}. Also, for the combination of mixed ($\eta-\mu$)-M{\'a}laga scenario, our demonstrated expressions in \eqref{a17} and \eqref{SOP_final} demonstrate similar result with \cite[Eqs.~(21) \& (26),]{yang2018physical} for the condition ($\alpha_m=\alpha_v=2$, $\mu_m=\mu_v>0$, and $\eta\geq0$).

\section{Numerical Results}

In this section, we provide graphical representations of our derived expressions for the performance metrics to demonstrate a better understanding on the effectiveness of the proposed scenario. MC simulations have also been performed in MATLAB for verifying the accuracy of the derived analytical expressions. Although we consider $\tilde{\alpha_r}=\tilde{\alpha_v}=\tilde{\alpha}$ in the expressions of the derived secrecy metrics, we adopt numerical approach to observe the impacts of $\tilde{\alpha_r}$ and $\tilde{\alpha_v}$ individually without any loss of generality. During numerical analysis, the values of $(\alpha_o,\beta_o)$ are considered as ($2.296, 2$), ($4.2, 3$), and ($8, 4$) for strong, moderate, and weak turbulence conditions, respectively \cite{trinh2016mixed}.

\begin{figure}[h!]
\vspace{-10mm}
    \centerline{\includegraphics[width=0.45\textwidth]{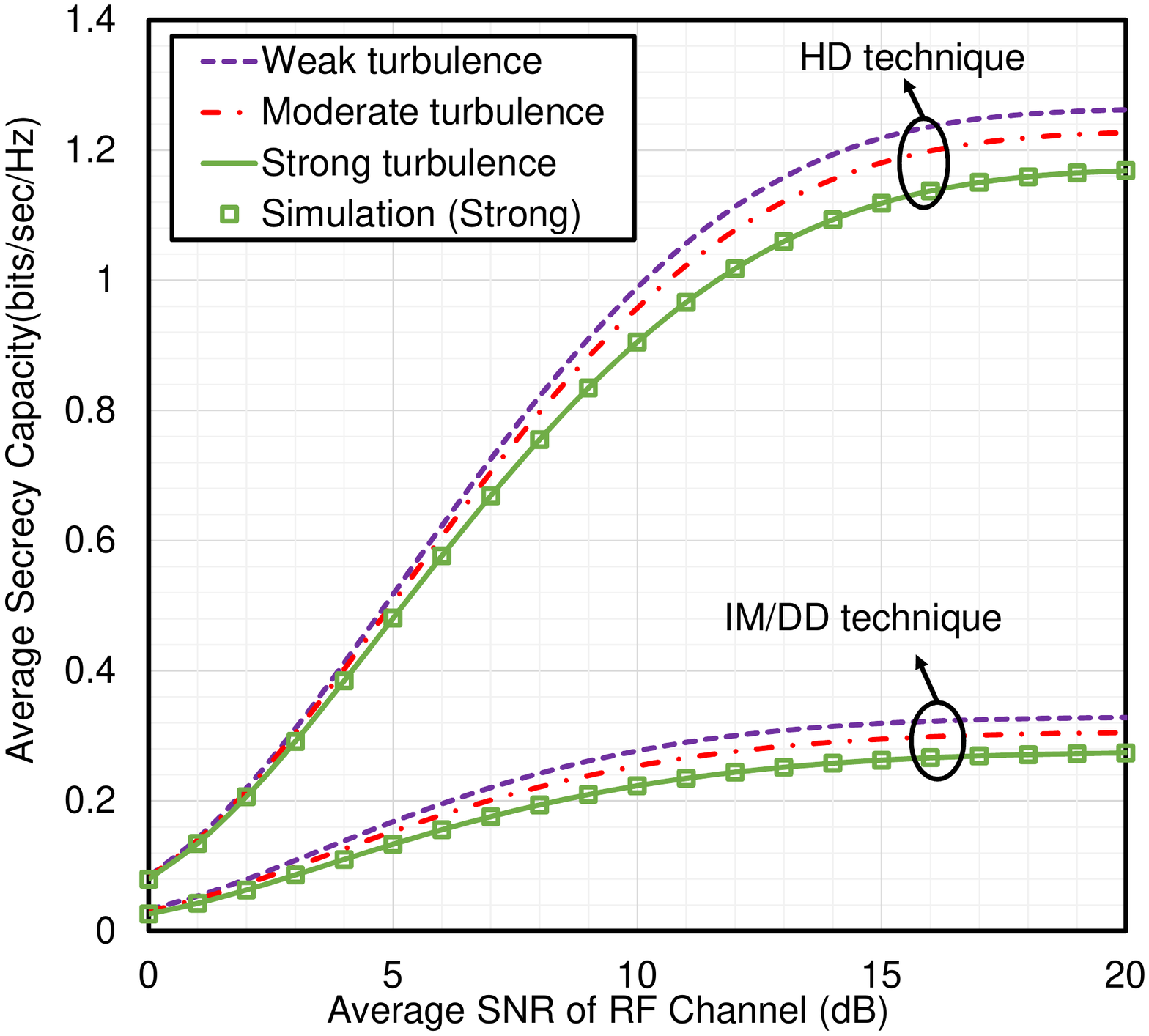}}
    \vspace{-25mm}
    \caption{The ASC versus ${\omega}_r$ for selected values of
    $\alpha_d, \beta_d$, and $r$ where $\alpha_r=\alpha_v=4$, $\mu_r=\mu_v=1$, $\eta_r=\eta_v{\approx}1$, $\Omega_d=g_d=2$, $u_r=10$ dB, and $\omega_v=0$ dB.}
    \label{fig:2}
    \vspace{-2mm}
\end{figure}

\begin{figure}[h!]
\vspace{-10mm}
    \centerline{\includegraphics[width=0.45\textwidth]{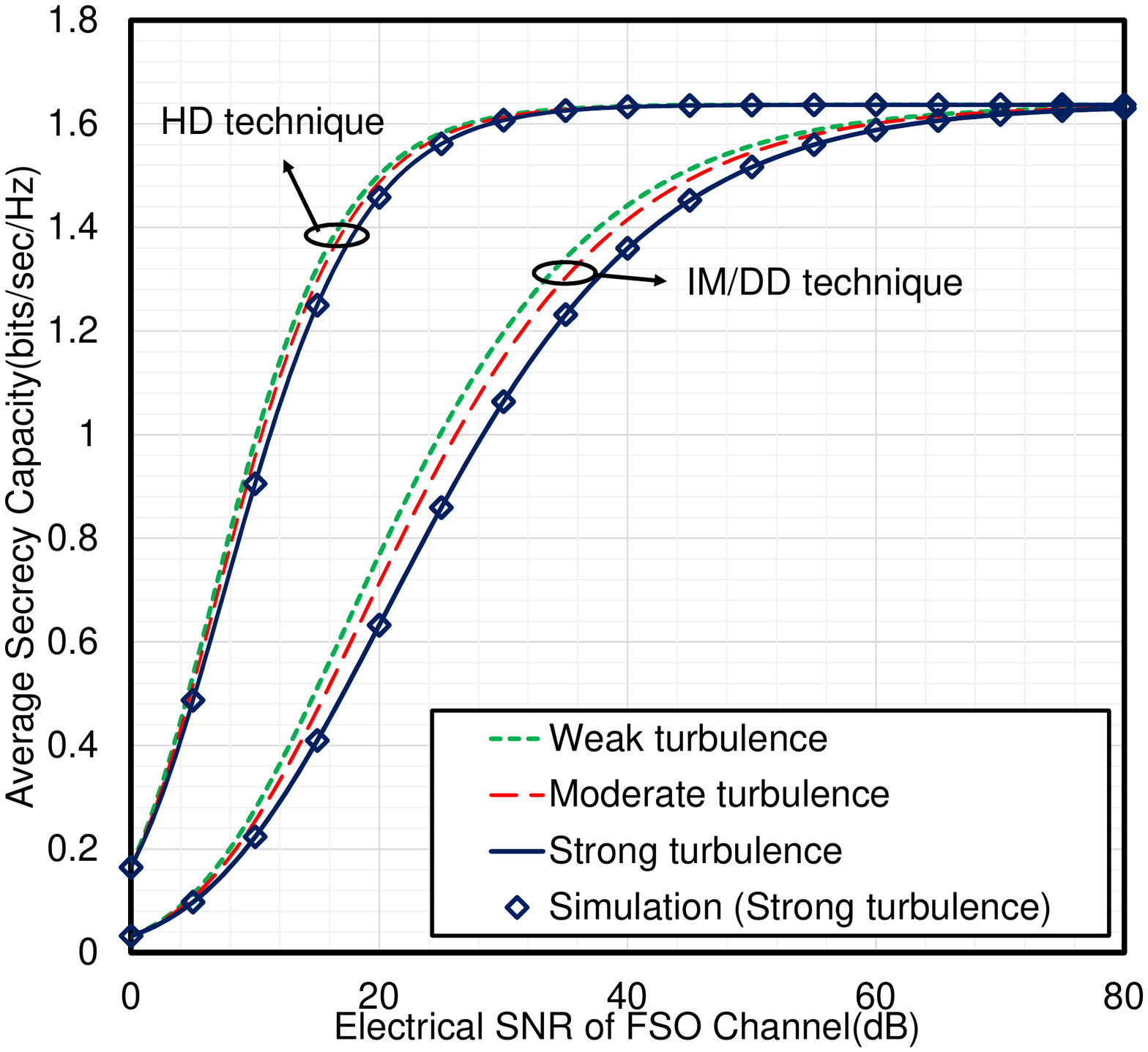}}
    \vspace{-25mm}
    \caption{The ASC versus $u_r$ for selected values of $\alpha_d$, $\beta_d$, and $r$ where $\alpha_r=\alpha_v=4$, $\mu_r=\mu_v=1$, $\eta_r = \eta_v \approx 1$, $\Omega_d=g_d=2$, $\omega_r=10$ dB, and $\omega_v=0$ dB.}
    \label{fig:3}
    \vspace{-2mm}
\end{figure}

\begin{figure}[h!]
\vspace{-10mm}
    \centerline{\includegraphics[width=0.45\textwidth]{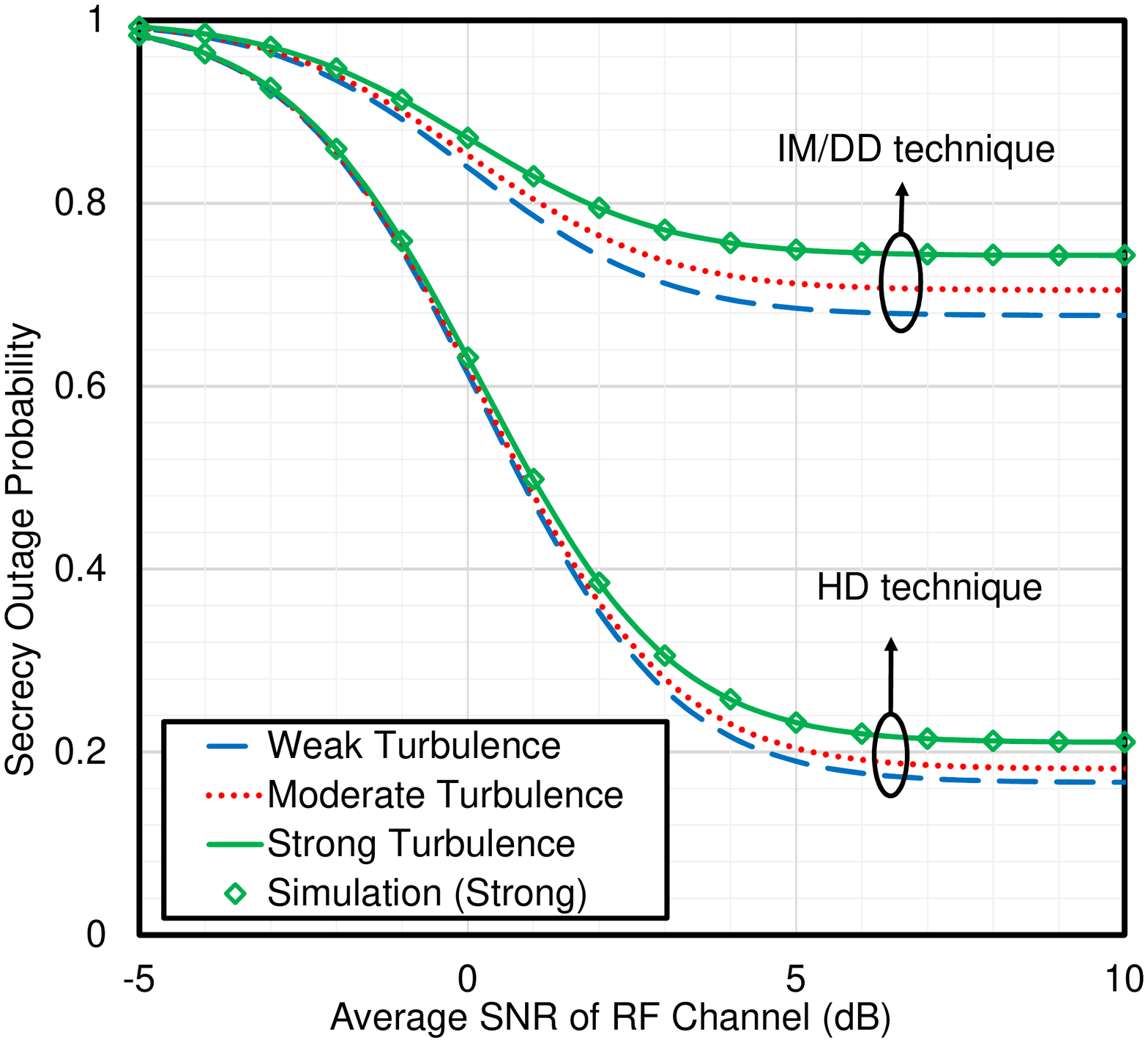}}
    \vspace{-25mm}
    \caption{The SOP versus $\omega_m$ for selected values of $\alpha_d$, $\beta_d$, and $r$ where $\Omega_d=g_d=2$, $\varepsilon=6.7$, $\alpha_r=\alpha_v=4$, $\mu_r=\mu_v=1$, $\eta_r = \eta_v \approx 1$, $R_s=0.1$ bits/sec/Hz, $\omega_v=0$ dB, and $u_r=10$ dB.}
    \label{fig:4}
    \vspace{-2mm}
\end{figure}

\begin{figure}[h!]
\vspace{-10mm}
    \centerline{\includegraphics[width=0.45\textwidth]{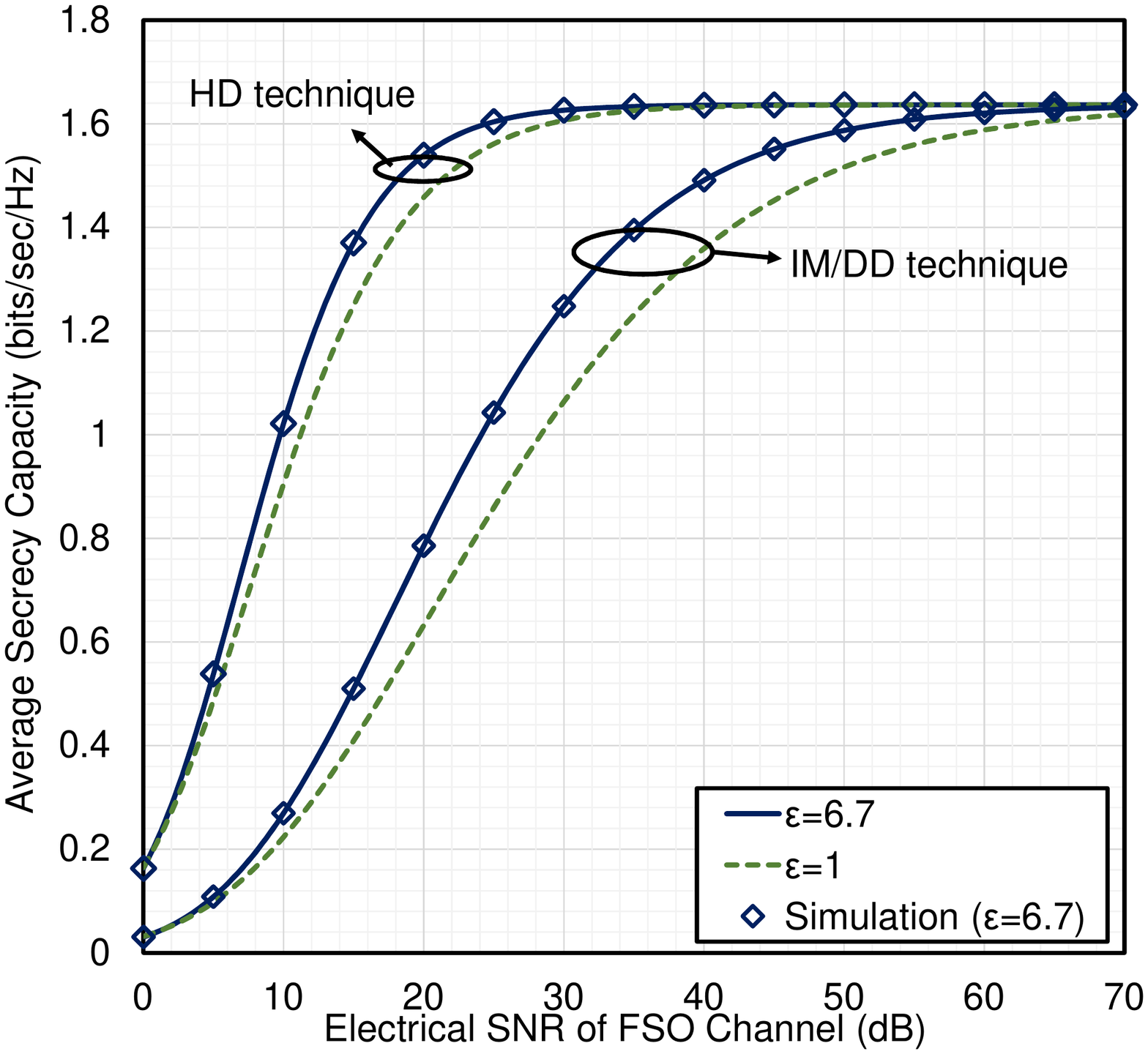}}
    \vspace{-25mm}
    \caption{The ASC versus $u_r$ for selected values of $\varepsilon$ and $r$ where $\alpha_d=2.296$, $\beta_d=2$, $\eta_r = \eta_v \approx 1$, $\Omega_d=g_d=2$, $\omega_r=10$ dB, and $\omega_v=0$ dB.}
    \label{fig:5}
    \vspace{-2mm}
\end{figure}

\begin{figure}[h!]
\vspace{-10mm}
    \centerline{\includegraphics[width=0.45\textwidth]{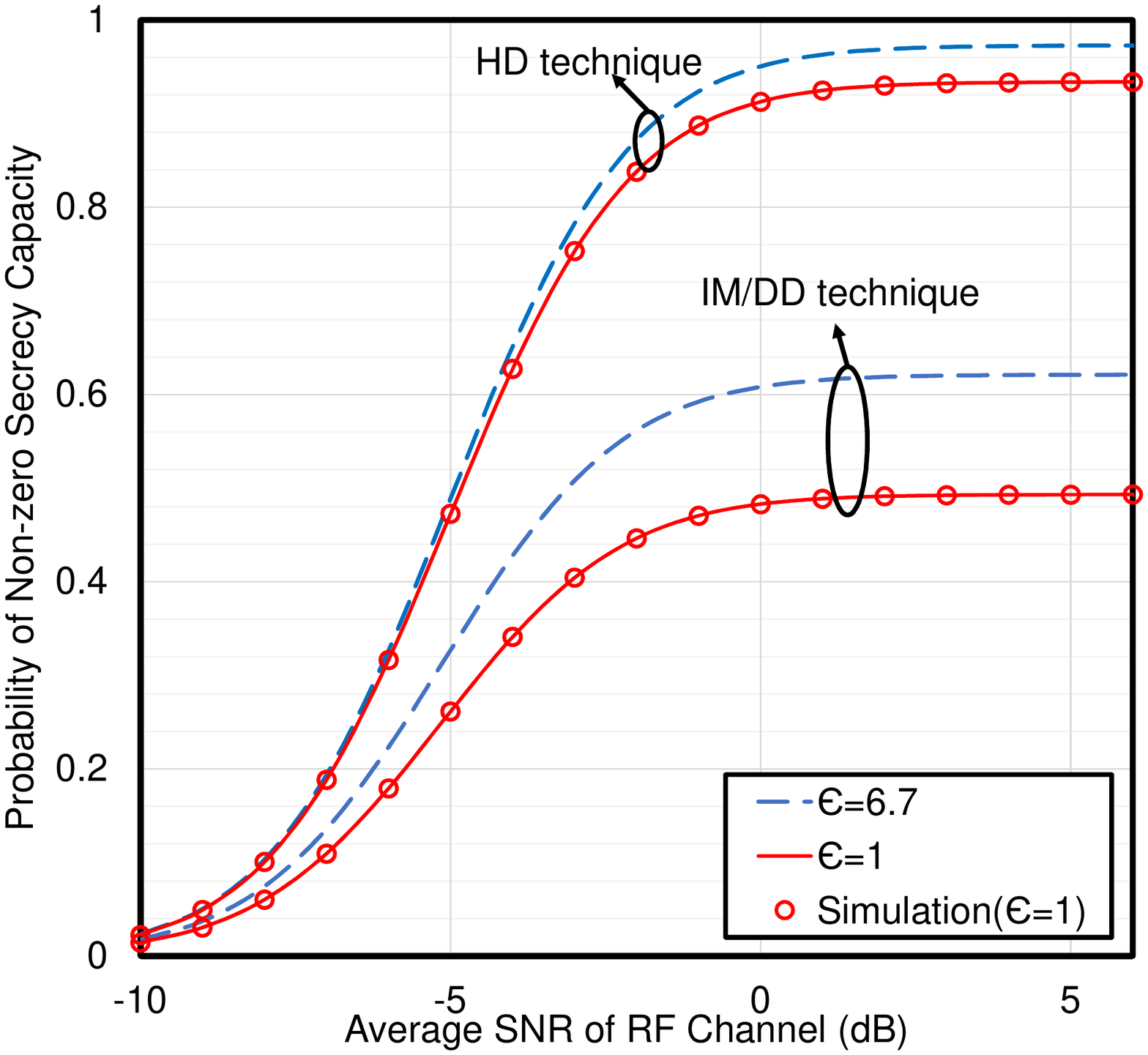}}
    \vspace{-25mm}
    \caption{The PNSC versus $\omega_r$ for the selected values of $\varepsilon$ and $r$ where $\Omega_d=g_d=2$, $\alpha_r=\alpha_v=4$, $\eta_r = \eta_v \approx 1$, $\mu_r=\mu_v=1$, $\omega_v=-5$ dB, and $u_r=10$ dB.}
    \label{fig:6}
    \vspace{-2mm}
\end{figure}

The comparison between the HD and IM/DD detection techniques in terms of security enhancement is graphically analyzed in Figs. \ref{fig:2}, \ref{fig:3}, \ref{fig:4}, \ref{fig:5}, and \ref{fig:6} against the performance metrics ASC, PNSC, and SOP. By simply observing these figures, we can conclude that the HD detection technique significantly outperforms the IM/DD technique as seen in \cite{saber2019physical,lei2018secrecy2,vellakudiyan2019performance,balti2018mixed,balti2017aggregate,palliyembil2018capacity}. This is because the IM/DD technique is based on the square of the intensity thereby excludes the effects of negative intensity. On the other hand, the HD technique considers the full scope of the intensity thereby providing better performance than the IM/DD.


The effects of atmospheric turbulence on the system performance are also demonstrated in Figs. \ref{fig:2}, \ref{fig:3}, and \ref{fig:4} where the system performance will be at its best for no or weak atmospheric turbulence condition and becomes vulnerable with the increasing amount of turbulence as proved in \cite{odeyemi2020secrecy,lei2018secrecy,pattanayak2020physical,hu2019performance,palliyembil2018capacity}. This characteristic is observed due to the fact that with increasing turbulence, the SNR at the receiver becomes smaller that in turn deteriorates the system security. The MC simulation demonstrates a close agreement with our analytical results, which point out the exactitude of the analytical expressions.

Pointing error signifies the misalignment between transmitter and receiver that causes huge data loss in FSO systems and it is an important factor that controls the integrity of FSO links. To evaluate the effect of the pointing error, the values of $\varepsilon$ have been appointed as $1$ and $6.7$ that corresponds to high and low pointing errors, respectively. As the FSO network depends on the LOS communication, the secrecy of the proposed system degrades with a higher pointing error between the transmitting and receiving aperture as demonstrated in Figs. \ref{fig:5} and \ref{fig:6}. Similar results are also achieved in \cite{zedini2014performance,odeyemi2018physical,islam2020secrecy} that justify our results.

\begin{figure}[h!]
\vspace{-10mm}
    \centerline{\includegraphics[width=0.45\textwidth]{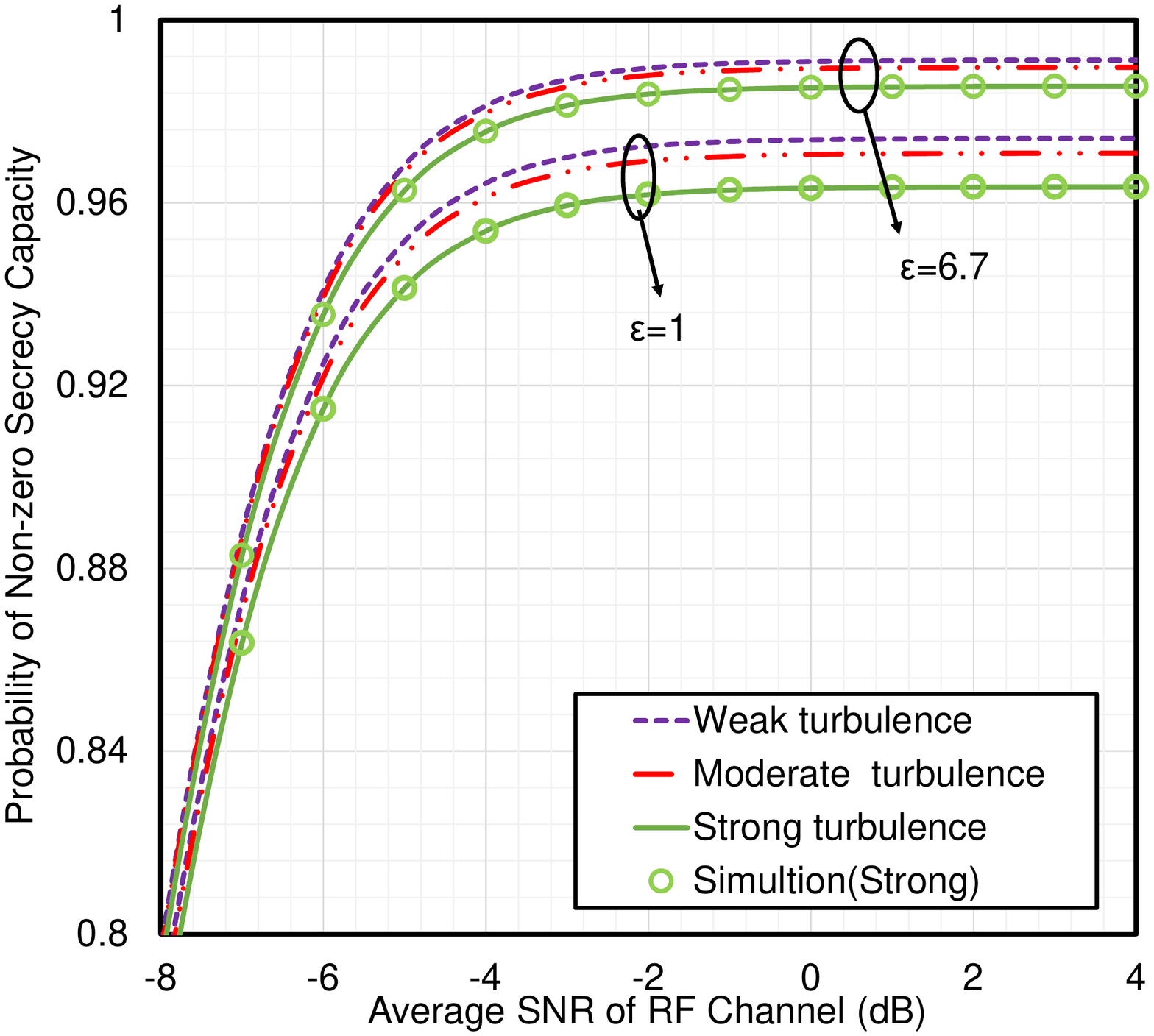}}
    \vspace{-25mm}
    \caption{The PNSC versus $\omega_r$ for the selected values of $\alpha_d$, $\beta_d$, and $r$ where $\Omega_d=g_d=2$, $\alpha_r=\alpha_v=4$, $\mu_r=\mu_v=1$, $\eta_r = \eta_v \approx 1$, $\omega_v=-10$ dB, and $u_r=10$ dB.}
    \label{fig:7}
    \vspace{-2mm}
\end{figure}

\begin{figure}[h!]
\vspace{-10mm}
    \centerline{\includegraphics[width=0.45\textwidth]{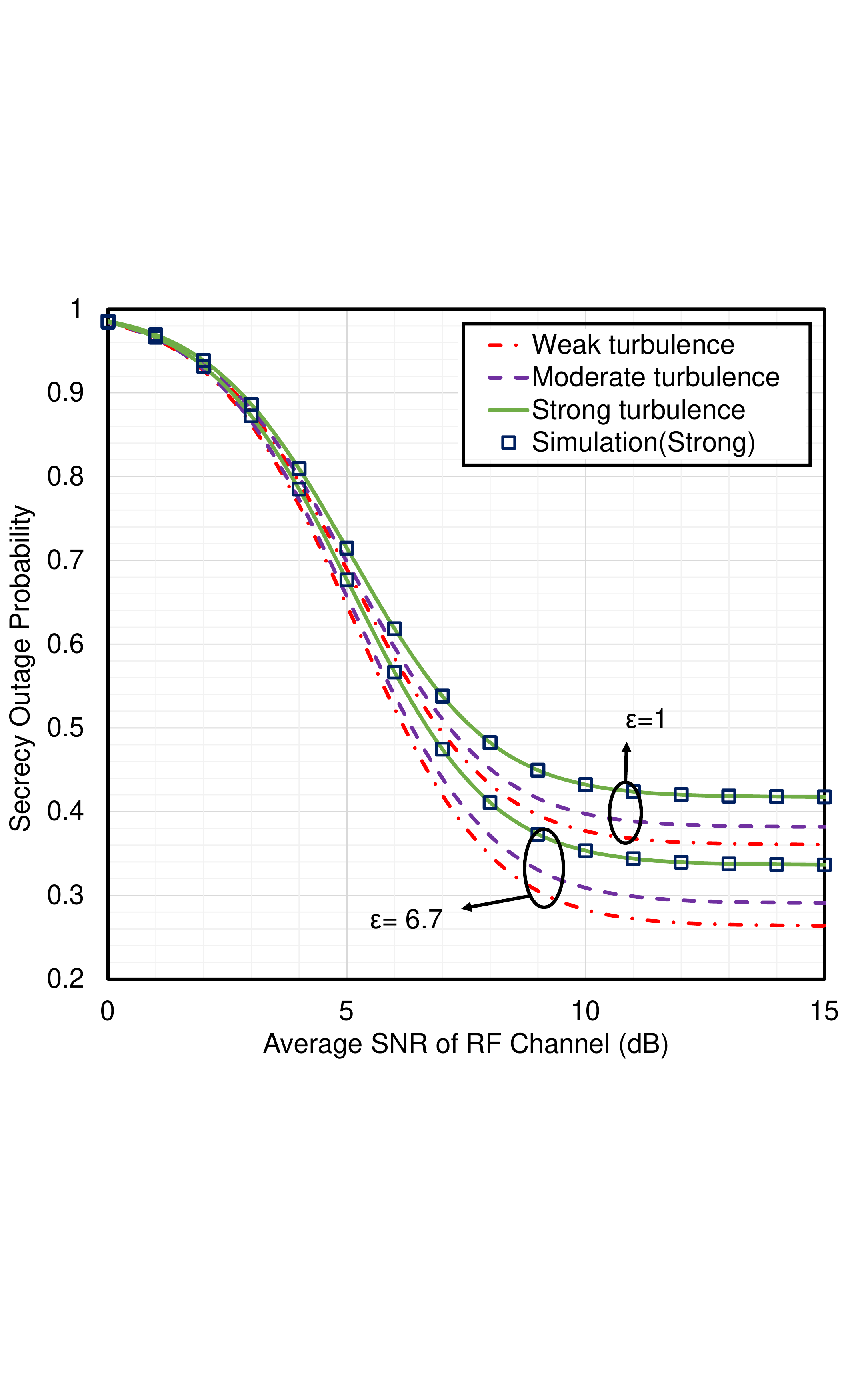}}
    \vspace{-25mm}
    \caption{The SOP versus $\omega_r$ for selected values of $\alpha_d$, $\beta_d$, and $r$ where $\Omega_d=2$, $\alpha_r=\alpha_v=4$, $\mu_r=\mu_v=1$, $\eta_r = \eta_v \approx 1$, $g_d=2$, $R_s=0.1$ bits/sec/Hz, and $\omega_v=5$ dB.}
    \label{fig:8}
    \vspace{-2mm}
\end{figure}

A comprehensive study on the effect of both pointing error and turbulence on the security of the proposed model is presented in Figs. \ref{fig:7} and \ref{fig:8} for PNSC and SOP, respectively. In both figures, the turbulence is varied for small and large pointing errors and it can be observed that the performance of the proposed system reduces at higher pointing error, as established in \cite{islam2020secrecy}. One can try to improve the system quality by decreasing the atmospheric turbulence but won't be able to remedy the situation completely, as analyzed in Figs. \ref{fig:7} and \ref{fig:8}, because the effect of pointing error is much higher than the turbulence created in the atmosphere. Hence, it is very important to maintain a lower pointing error for secure RF-FSO communications, regardless of the atmospheric turbulence conditions.

\begin{figure}[h!]
\vspace{-10mm}
    \centerline{\includegraphics[width=0.45\textwidth]{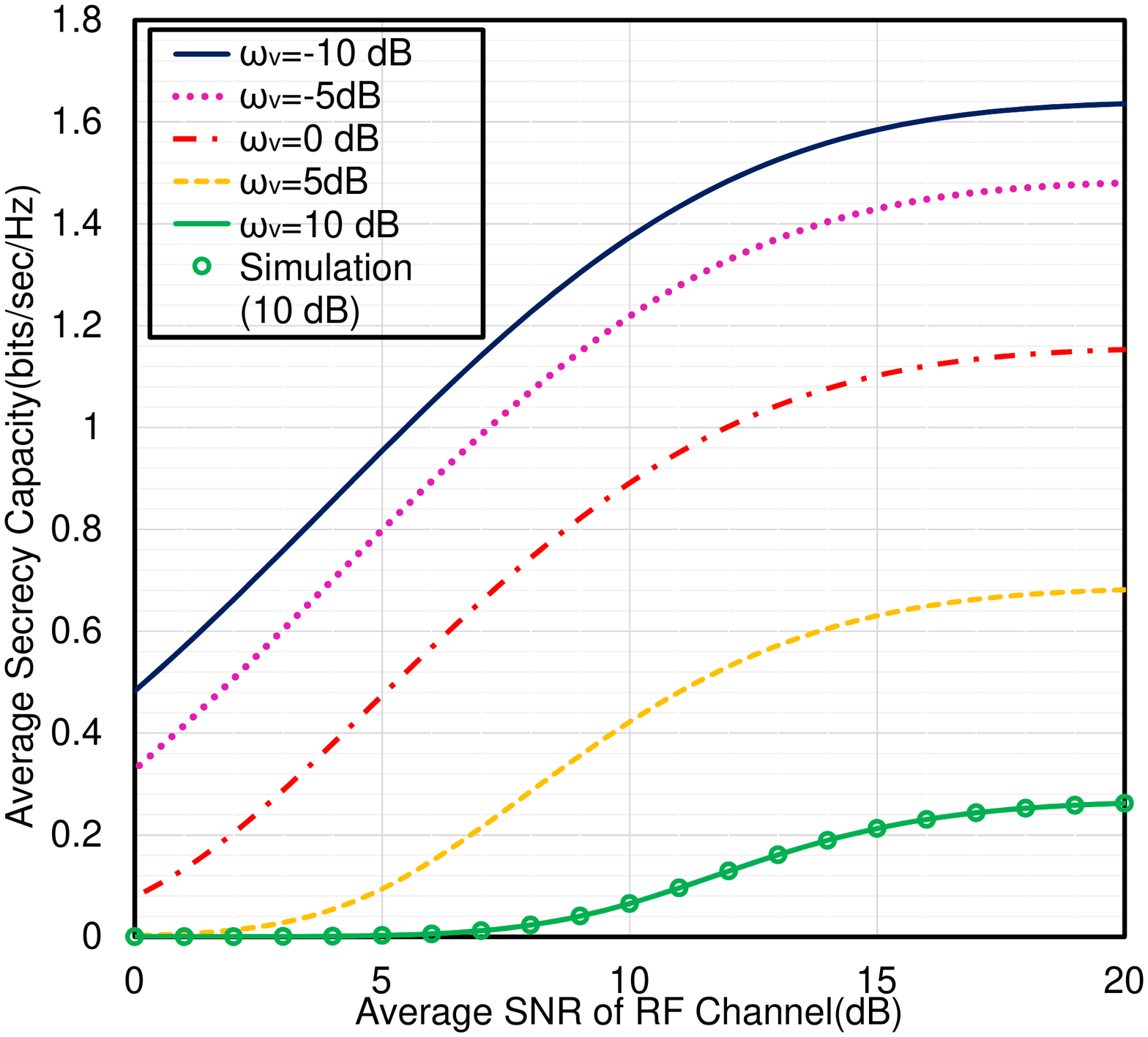}}
    \vspace{-25mm}
    \caption{The ASC versus ${\omega}_r$ for selected values of $\omega_v$ where $\alpha_r=\alpha_v=4$, $\mu_r=\mu_v=1$, $\eta_r = \eta_v \approx 1$, $\alpha_d=2.296$, $\beta_d=2$, $r=1$, $\Omega_d=g_d=2$, and $u_r=10$ dB.}
    \label{fig:9}
    \vspace{-2mm}
\end{figure}

Figure \ref{fig:9} demonstrates the effect of average SNR of the eavesdropper channel ($\omega_{v}$) plotted against $\omega_{r}$ on the ASC of this proposed work. It is clearly illustrated from the Fig. that with increasing values of $\omega_{v}$, a sharp decay occurs in the secrecy capacity of the system. The reason behind this is the increment in $\omega_{v}$ actually increases the quality of the eavesdropper channel and thus the eavesdropper gains unauthorized access to more data that compromises the secrecy of the transmitted data. The authors in \cite{islam2020secrecy, badrudduza2020enhancing} have also deduced similar results.

\begin{figure}[h!]
\vspace{-10mm}
    \centerline{\includegraphics[width=0.45\textwidth]{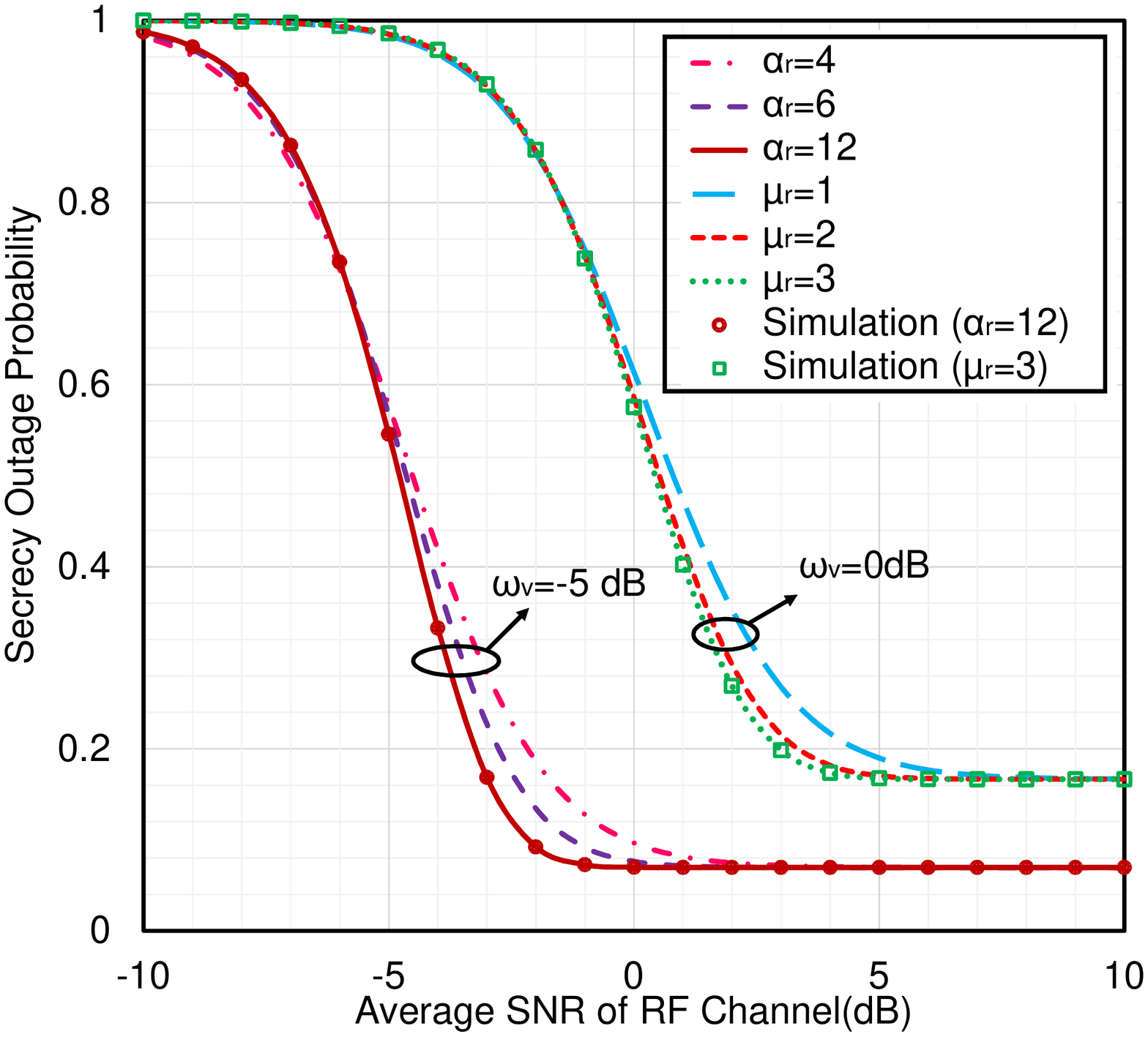}}
    \vspace{-25mm}
    \caption{The SOP versus $\omega_r$ for selected values of $\alpha_r$, $\mu_r$, and $\omega_v$ where $\Omega_d=g_d=2$, $\alpha_v=4$, $\mu_v=1$, $\eta_r = \eta_v \approx 1$, $\varepsilon=1$, $R_s=0.1$ bits/sec/Hz, and $u_r=10$ dB.}
    \label{fig:10}
    \vspace{-2mm}
\end{figure}

\begin{figure}[h!]
\vspace{-10mm}
    \centerline{\includegraphics[width=0.50\textwidth]{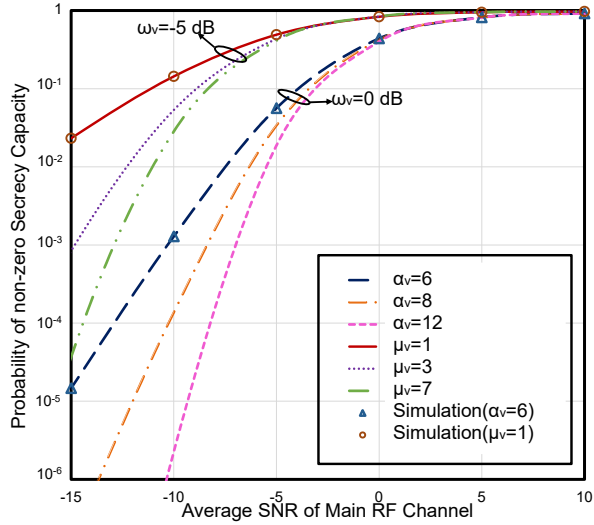}}
    \vspace{-25mm}
    \caption{The PNSC versus $\omega_r$ for selected values of $\alpha_v$, $\mu_v$, and $\omega_v$ where $\Omega_d=g_d=2$, $\alpha_r=2$, $\mu_r=1$, $\eta_r = \eta_v \approx 1$, $\varepsilon=1$, $\alpha_d=8$, $\beta_d=4$, and $u_r=15$ dB.}
    \label{fig:11}
    \vspace{-2mm}
\end{figure}

The effect of non-linearity along with the number of multipath clusters has a tremendous effect on the environmental fading through which the RF signal has to propagate towards the destination. The influence of these parameters on the proposed system is demonstrated in Figs. \ref{fig:10} and \ref{fig:11} considering variations in main channel and eavesdropper channel, respectively. Fig. \ref{fig:10} illustrates the security can be enhanced by increasing the value of $\alpha_r$ and $\mu_r$. The increment in $\alpha_r$ reduces the total non-linearity of the system that in turn reduces the environmental fading. On the other hand, with increasing values of $\mu_r$, total multipath clusters are increased that increases the possibility of achieving a better SNR at the relay, and thus security is increased. Similar analysis is displayed in Fig. \ref{fig:11} where these parameters for the eavesdropper channel are varied. We can observe that the security of the system reduces with increasing $\alpha_v$ and $\mu_v$ because their increment actually consolidates the eavesdropper channel that becomes able to steal more data from the legitimate RF channel. Same characteristics are also perceived in \cite{salameh2019end,badarneh2015performance,sarker2020secrecy,gupta2018performance}. We can observe that a SOP floor is reached in Fig. \ref{fig:10} that is due to the limitation of RF link rather than the FSO link.
\begin{table*}[h!]
\centering
\caption{Generalization of the Proposed Model.}
\label{tab:abc}
\begin{tabular}{|c|c|}
\hline
Envelope distribution   & Parameters                       \\ \hline \hline
($\alpha-\mu$)-M{\'a}laga       & $\alpha_r=\alpha_v=4$, $\mu_r=\mu_v=1$, $g_d=2$, $\Omega_d=1$, $\rho=0$ \\ \hline
($\eta-\mu$)-M{\'a}laga          & $\alpha_r=\alpha_v=6$, $\mu_r=\mu_v=3$, $g_d=2$, $\Omega_d=4$, $\rho=0$ \\ \hline
(Nakagami-$m$)-($\Gamma-\Gamma$) & $\alpha_r=\alpha_v=2$, $\mu_r=\mu_v=1$,  $g_d=0$, $\Omega_d=1$, $\rho=1$ \\ \hline
Rayeleigh-($\Gamma-\Gamma$)  & $\alpha_r=\alpha_v=2$, $\mu_r=\mu_v=0.5$, $g_d=0$, $\Omega_d=1$, $\rho=1$ \\ \hline
Weibull-Lognormal      & $\alpha_r=\alpha_v=4$, $\mu_r=\mu_v=0.5$,  $g_d=0.0001$, $\Omega_d=1.3265$, $\rho=0$ \\ \hline
\end{tabular}
\end{table*}

\begin{figure}[h!]
\vspace{-10mm}
    \centerline{\includegraphics[width=0.45\textwidth]{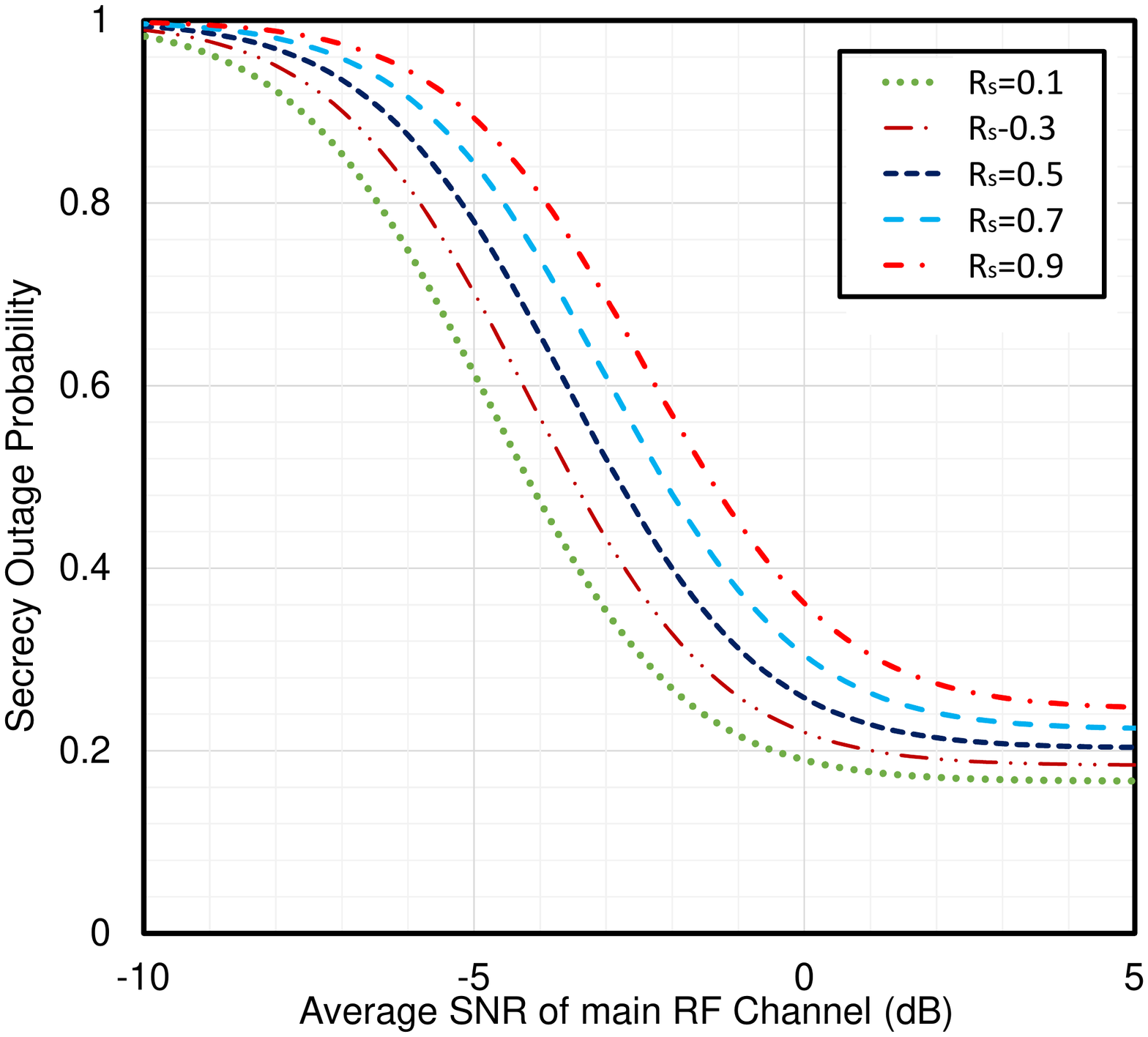}}
    \vspace{-25mm}
    \caption{The SOP versus ${\omega}_r$ for selected values of
    $R_s$, where $\alpha_r=\alpha_v=4$, $\mu_r=\mu_v=1$, $\eta_r = \eta_v \approx 1$, $\alpha_d=8$, $\beta_d=4$, $r=1$, $\Omega_d=g_d=2$, $u_r=5$ dB, and $\omega_v=-5$ dB.}
    \label{fig:12}
    \vspace{-2mm}
\end{figure}

The target secrecy rate ($R_{s}$), as defined earlier, is a threshold capacity that determines whether the instantaneous secrecy capacity of the transmitting signal is below or greater than this rate. For supreme secrecy, the instantaneous secrecy rate must be higher than the $R_{s}$. But with increasing $R_{s}$, the probability of the instantaneous secrecy capacity being higher than the $R_{s}$ becomes smaller. As a result, the secrecy outage of the system increases. This phenomenon is exhibited in Fig. \ref{fig:12} where SOP of the proposed system is plotted against ${\omega}_r$ for varying values of $R_{s}$. This same behaviour has also been observed in \cite{salameh2019end, badrudduza2021security}.

\quad

\noindent
\textbf{Special Cases of the Proposed Scenario:}

\begin{figure}[h!]
\vspace{-10mm}
    \centerline{\includegraphics[width=0.50\textwidth]{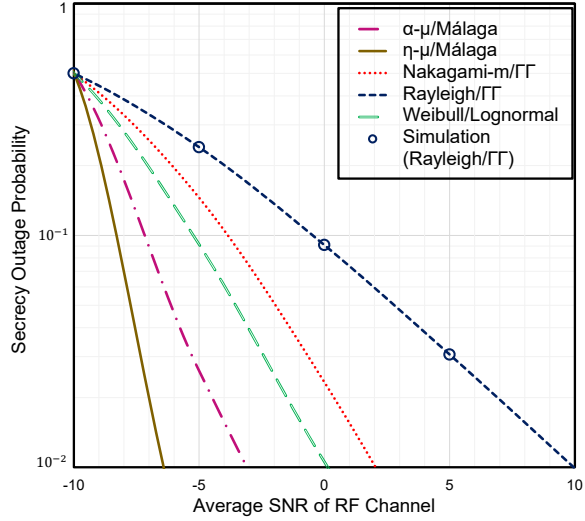}}
    \vspace{-25mm}
    \caption{The SOP versus ${\omega}_r$ for selected values of $\alpha_r$, $\alpha_v$, $\mu_r$, $\mu_v$, $\rho$, $g_d$, and $\Omega_d$ where $\alpha_d=8$, $\beta_d=4$, $r=1$, $\eta_r = \eta_v \approx 1$, $u_r=20$ dB, and $\omega_v=-10$ dB.}
    \label{fig:13}
    \vspace{-2mm}
\end{figure}

As discussed earlier, the communications system proposed in this work considers generalized fading conditions at both RF and FSO link that enables this model to be applicable for a wide variety of fading scenarios. Figure \ref{fig:13} demonstrates a few well-known combinations of RF-FSO links that have been studied in some earlier works. Hence, by altering some parameter values (as given in Table \ref{tab:abc}) in our proposed work, we can analyze all these models that are previously studied in \cite{lei2017secrecy, yang2018physical,lei2018secrecy,abd2016security} (considering the single antenna in \cite{lei2018secrecy,abd2016security}) as well as many other combinations of RF-FSO hops. This prominent characteristic makes our analysis more unique and novel relative to others.

\section{Conclusion}
This work deals with the mathematical modeling of a secure RF-FSO network considering a variable gain relaying scheme for $\alpha-\eta-\mu$ fading model at the RF hop and M\'alaga model at the FSO hop. The secrecy analysis of the proposed scenario has been performed via deriving the closed-form expressions for the performance metrics i.e. ASC, lower bound of SOP, and PNSC incorporating the adverse effects of the fading, pointing errors, and atmospheric turbulence for both HD and IM/DD techniques. We further corroborate the analytical results via MC simulations. As predicted, HD demonstrates superior secrecy performance than the IM/DD technique. Although both pointing errors and atmospheric turbulence are the reasons behind the degradation of the security, the effect of pointing error is far worse than the turbulence. Moreover, we focus on the generalization of the existing works incorporating more complicated composite channels such as $\alpha-\eta-\mu$ fading and unified M\'alaga turbulent models that will pave the way for the communication engineers to design more practical secure digital communication scenarios via exploiting the randomness of the wireless channels.
\appendix
\section*{Appendix}
\section{Proof of ASC}
\label{ASC}
For mathematical simplification, we have considered $\tilde{\alpha_r}=\tilde{\alpha_v}=\tilde{\alpha}$. The final expressions of $\mathcal{S}_1$, $\mathcal{S}_2$, $\mathcal{S}_3$, and $\mathcal{S}_4$ are given below.

\vspace{3mm}
\noindent
\textbf{Calculation of $\mathcal{S}_1$:} 

Using the identities \cite[Eqs.~(8.4.3.1), (8.4.2.5),]{prudnikov1988integrals}, \eqref{19a} can be given as
\begin{align}
\label{20}
    \mathcal{S}_1= \int_0^\infty \gamma^{\tilde{\alpha} t_1} G_{1,1}^{1,1}\left[\gamma \biggl|
    \begin{array}{c}
        0 \\
        0 \\
    \end{array}
    \right] G_{0,1}^{1,0}\left[u_{2}\gamma^{\tilde{\alpha}} \biggl|
    \begin{array}{c}
        . \\
        0 \\
    \end{array}
    \right] d\gamma.
\end{align}
Now, applying \cite[Eqs.~(2.24.1.1),]{prudnikov1988integrals} in \eqref{20} and performing the integration, we obtain
\begin{align}
    \label{21}
    \mathcal{S}_1=(2 \pi)^{1-\tilde{\alpha}}G_{\tilde{\alpha},1+\tilde{\alpha}}^{1+\tilde{\alpha},\tilde{\alpha}}\left[u_{2} \biggl|
    \begin{array}{c}
       \Delta(\tilde{\alpha}, -\tilde{\alpha}t_1) \\
        0,\Delta(\tilde{\alpha}, -\tilde{\alpha}t_1) \\
    \end{array}
    \right],
\end{align}
where $\Delta(a,b)=\frac{b}{a}, \frac{b+1}{a}, ..., \frac{b+a-1}{a}$.

\vspace{3mm}
\noindent
\textbf{Calculation of $\mathcal{S}_2$:}

We derive $\mathcal{S}_2$ by following the similar process as $\mathcal{S}_1$ and $\mathcal{S}_2$ can be expressed as
\begin{align}
    \nonumber
    \mathcal{S}_2&= \int_0^\infty \gamma^{\tilde{\alpha} \mathcal{T}}  G_{1,1}^{1,1}\left[\gamma \biggl|
    \begin{array}{c}
    0 \\
    0 \\
    \end{array}
    \right] G_{0,1}^{1,0}\left[\Im \gamma^{\tilde{\alpha}} \biggl|
    \begin{array}{c}
    . \\
    0 \\
    \end{array}
    \right] d\gamma
    \\
    &=(2 \pi)^{1-\tilde{\alpha}}G_{\tilde{\alpha},1+\tilde{\alpha}}^{1+\tilde{\alpha},\tilde{\alpha}}\left[\Im \biggl|
    \begin{array}{c}
       \Delta(\tilde{\alpha}, -\tilde{\alpha}\mathcal{T}) \\
        0,\Delta(\tilde{\alpha}, -\tilde{\alpha}\mathcal{T}) \\
    \end{array}
    \right],
\end{align}
where $\mathcal{T}=t_1+t_2$ and $\Im=u_{2}+q_2$.

\vspace{3mm}
\noindent
\textbf{Calculation of $\mathcal{S}_3$:}

Utilizing the identities \cite[Eqs.~(8.4.3.1), (8.4.2.5),]{prudnikov1988integrals} in \eqref{19c}, $\mathcal{S}_3$ can be expressed as
\begin{align}
    \label{23}
    \mathcal{S}_3&= \int_0^\infty \gamma^{\tilde{\alpha} t_1} G_{1,1}^{1,1}\left[\gamma \biggl|
    \begin{array}{c}
        0 \\
        0 \\
    \end{array}
    \right] G_{0,1}^{1,0}\left[u_{2}\gamma^{\tilde{\alpha}}  \biggl|
    \begin{array}{c}
        . \\
        0 \\
    \end{array}
    \right] 
 G_{r+1,3r+1}^{3r,1}\left[\frac{F \gamma}{u _r} \biggl|
    \begin{array}{c}
        1,l_1 \\
        l_2,0 \\
    \end{array} 
    \right]d\gamma.
    \end{align}
For deriving the closed form expression of $\mathcal{S}_3$, we must integrate \eqref{23} within the limit $0$ to $\infty$ that is mathematically intractable. So, for obtaining $\mathcal{S}_3$ in closed-form, Meijer's $G$ function terms are converted into Fox's $H$ functions utilizing \cite[Eqs.~(8.3.2.21),]{prudnikov1988integrals} as
    \begin{align}
    \label{24}
    \mathcal{S}_3&=\int_0^\infty \gamma^{\tilde{\alpha} t_1} H_{1,1}^{1,1}\left[\gamma \biggl|
    \begin{array}{c}
         {[}0,1{]}\\
         {[}0,1{]} \\
    \end{array}
    \right] H_{0,1}^{1,0}\left[u_{2}\gamma^{\tilde{\alpha}} \biggl|
    \begin{array}{c}
        . \\
         {[}0,1{]} \\
    \end{array}
    \right]
    H_{r+1,3r+1}^{3r,1}\left[\frac{F \gamma}{u _r} \biggl|
    \begin{array}{c}
         {[}1,1{]},{[}l_1,1{]} \\
        {[}l_2,1{]}, {[}0,1{]} \\
    \end{array} 
    \right]d\gamma.
    \end{align}
Again, for mathematical simplification, we assume $x=\gamma^{\tilde{\alpha}}$. So, \eqref{24} can be expressed as
    \begin{align}
   \label{25}
    \mathcal{S}_3&=\frac{1}{\tilde{\alpha}}\int_0^\infty x^{  \mathcal{M}_1-1} H_{0,1}^{1,0}\left[u_{2}x \biggl|
    \begin{array}{c}
        . \\
         {[}0,1{]} \\
    \end{array}
    \right] H_{1,1}^{1,1}\left[x^{\frac{1}{\tilde{\alpha}}} \biggl|
    \begin{array}{c}
         {[}0,1{]}\\
         {[}0,1{]} \\
    \end{array}
    \right] 
    H_{r+1,3r+1}^{3r,1}\left[\frac{F x^{\frac{1}{\tilde{\alpha}}}}{u _r} \biggl|
    \begin{array}{c}
         {[}1,1{]},{[}l_1,1{]} \\
        {[}l_2,1{]}, {[}0,1{]} \\
    \end{array} 
    \right]dx.
    \end{align}
Now utilizing \cite[Eq.~(2.3),]{mittal1972integral},  \cite[Eq.~(3),]{lei2017secrecy}, the final expression of $\mathcal{S}_3$ is derived as
    \begin{align}
    \label{26}
    \mathcal{S}_3&=\frac{1}{\tilde{\alpha} {u_{2}}^{\mathcal{M}_1}}   H_{1,0;1,1;r+1,3r+1}^{1,0;1,1;3r,1}\left[
    \begin{array}{c} 
     {[}1-\mathcal{M}_1; \frac{1}{\tilde{\alpha}}  , \frac{1}{\tilde{\alpha}}{]}   \\ 
     - \\ 
    \end{array} \biggl|
    \begin{array}{c}
    {[}0,1{]} \\
    {[}0,1{]} \\
    \end{array}
    \biggl|
    \begin{array}{c}
    {[}1,1{]},{[}l_1,1{]} \\
        {[}l_2,1{]}, {[}0,1{]} \\
    \end{array}
    \biggl|\frac{1}{{u_2}^{\frac{1}{\tilde{\alpha}}}}, \frac{F}{u_r {u_2}^{\frac{1}{\tilde{\alpha}}}}
    \right],
\end{align}
where $\mathcal{M}_1=\frac{1}{\tilde{\alpha}}+t_1$, $H_{m1,n1:m2,n2:m3,q3}^{p1,q1:p2,q2:p3,q3}[.]$ is the extended generalized bivariate Fox's $H$ function.

\vspace{3mm}
\noindent
\textbf{Calculation of $\mathcal{S}_4$:}

Similar to $\mathcal{S}_3$, $\mathcal{S}_4$ is obtained as
\begin{align}
    \nonumber
    \mathcal{S}_4&= \int_0^\infty \gamma^{\tilde{\alpha} \mathcal{T}}  G_{1,1}^{1,1}\left[\gamma \biggl|
    \begin{array}{c}
    0 \\
    0 \\
    \end{array}
    \right] G_{0,1}^{1,0}\left[\Im \gamma^{\tilde{\alpha}} \biggl|
    \begin{array}{c}
    . \\
    0 \\
    \end{array}
    \right] 
     G_{r+1,3r+1}^{3r,1}\left[\frac{F \gamma}{u _r} \biggl|
    \begin{array}{c}
        1,l_1 \\
        l_2,0 \\
    \end{array} 
    \right]d\gamma
    \\
    \nonumber
    &=\int_0^\infty \gamma^{\tilde{\alpha}\mathcal{T}} H_{1,1}^{1,1}\left[\gamma \biggl|
    \begin{array}{c}
         {[}0,1{]}\\
         {[}0,1{]} \\
    \end{array}
    \right] H_{0,1}^{1,0}\left[\Im \gamma^{\tilde{\alpha}} \biggl|
    \begin{array}{c}
        . \\
         {[}0,1{]} \\
    \end{array}
    \right]
     H_{r+1,3r+1}^{3r,1}\left[\frac{F \gamma}{u _r} \biggl|
    \begin{array}{c}
         {[}1,1{]},{[}l_1,1{]} \\
        {[}l_2,1{]}, {[}0,1{]} \\
    \end{array} 
    \right]d\gamma
    \\
   \nonumber
    &=\frac{1}{\tilde{\alpha}}\int_0^\infty x^{  \mathcal{M}_2-1} H_{0,1}^{1,0}\left[\Im x \biggl|
    \begin{array}{c}
        . \\
         {[}0,1{]} \\
    \end{array}
    \right] H_{1,1}^{1,1}\left[x^{\frac{1}{\tilde{\alpha}}} \biggl|
    \begin{array}{c}
         {[}0,1{]}\\
         {[}0,1{]} \\
    \end{array}
    \right] 
    \times H_{r+1,3r+1}^{3r,1}\left[\frac{F x^{\frac{1}{\tilde{\alpha}}}}{u _r} \biggl|
    \begin{array}{c}
         {[}1,1{]},{[}l_1,1{]} \\
        {[}l_2,1{]}, {[}0,1{]} \\
    \end{array} 
    \right]dx
    \\
    &=\frac{1}{\tilde{\alpha} {\Im}^{\mathcal{M}_2}}   H_{1,0;1,1;r+1,3r+1}^{1,0;1,1;3r,1}\left[
    \begin{array}{c} 
     {[}1-\mathcal{M}_2; \frac{1}{\tilde{\alpha}}  , \frac{1}{\tilde{\alpha}}{]}   \\ 
     - \\ 
    \end{array} \biggl|
    \begin{array}{c}
    {[}0,1{]} \\
    {[}0,1{]} \\
    \end{array}
    \biggl|
    \begin{array}{c}
    {[}1,1{]},{[}l_1,1{]} \\
        {[}l_2,1{]}, {[}0,1{]} \\
    \end{array}
    \biggl|
    \biggl|\frac{1}{{\Im}^{\frac{1}{\tilde{\alpha}}}}, \frac{F}{u_r {\Im}^{\frac{1}{\tilde{\alpha}}}}
    \right],
\end{align}
where $\mathcal{M}_2=\frac{1}{\tilde{\alpha}}+\mathcal{T}$.

\section{Proof of SOP}
\label{SOP}

Plugging \eqref{a11} and \eqref{a12} into \eqref{a27} leads to
\begin{align}
    \nonumber
    P_{out}^L(R_s)&=1-\sum_{N_1=0}^{\infty} \sum_{N_2=0}^{\infty}\sum_{t_1=0  }^{W_1-1}u_{4} q_2 \theta^{\tilde{\alpha_r} t_1}\int_0^\infty e^{-(u_{2}\theta\gamma^{\tilde{\alpha_r}}+q_2\gamma^{\tilde{\alpha_v}})}
    \\
    &\times {\gamma}^{q_3+\tilde{\alpha_r}t_1} \biggl(1-\sigma \sum _{\tilde {m}_d=1}^{\beta_d } c_d G_{r+1,3r+1}^{3r,1}\left[\frac{F }{u _r}(\theta\gamma) \biggl|
    \begin{array}{c}
     1,l_1 \\
     l_2,0 \\
    \end{array}
    \right]\biggl)d\gamma.
\end{align}
Here, for mathematical tractability, we consider $\tilde{\alpha_r}=\tilde{\alpha_v}=\tilde{\alpha}$.

\vspace{3mm}
\noindent
\textbf{Calculation of $\mathcal{H}_{1}$:}

Using \cite[Eq.~(3.326.2),]{gradshteyn2014table}, $\mathcal{H}_{1}$ is expressed as
\begin{align}
    \mathcal{H}_{1}&=\int_0^{\infty}{\gamma}^{q_3+\tilde{\alpha}t_1}e^{-\kappa\gamma^{\tilde{\alpha}}}d\gamma={\frac{\Gamma(\mathcal{Z}_1)}{ \tilde{\alpha}{\kappa}^{\mathcal{Z}_1}}},
\end{align}
where $\mathcal{Z}_1=\frac{q_3+\tilde{\alpha_r}t_1+1}{\tilde{\alpha}}$ and $\kappa=u_{2}\theta+q_2$.

\vspace{3mm}
\noindent
\textbf{Calculation of $\mathcal{H}_{2}$:}

$\mathcal{H}_{2}$ in \eqref{SOP_final} is given as
\begin{align}
    \mathcal{H}_{2}=\int_0^{\infty} {\gamma}^{q_3+\tilde{\alpha}t_1} e^{-\kappa\gamma^{\tilde{\alpha}}} G_{r+1,3r+1}^{3r,1}\left[\frac{F }{u _r}(\theta\gamma) \biggl|
    \begin{array}{c}
     1,l_1 \\
     l_2,0 \\
    \end{array}
    \right] d\gamma.
\end{align}
Letting $I=\gamma^{\tilde{\alpha}}$ and utilizing \cite[Eqs.~(8.4.3.1) and (2.24.1.1),]{prudnikov1988integrals}, $\mathcal{H}_{2}$ can be expressed as
\begin{align}
    \nonumber
    &\mathcal{H}_{2}  ={\frac{1}{\tilde{\alpha}}} \int_0^{\infty} I^{\mathcal{Z}_1-1} e^{-\kappa I} G_{r+1,3r+1}^{3r,1}\left[\frac{F\theta }{u _r}I^{\frac{1}{\tilde{\alpha}}}\biggl|
    \begin{array}{c}
     1,l_1 \\
     l_2,0 \\
    \end{array}
    \right] dI
    \\
    &={\frac{(2\pi)^{(1-\tilde{\alpha})r}}{{\kappa}^{\mathcal{Z}_1}\tilde{\alpha}^{1-\mathcal{Z}_2}} } G_{\tilde{\alpha}r+\tilde{\alpha}+1,3\tilde{\alpha}r+\tilde{\alpha}}^{3\tilde{\alpha}r,\tilde{\alpha}+1}\left[\frac{(F\theta)^{\tilde{\alpha} } }{{u_r}^{\tilde{\alpha}} \kappa {\tilde{\alpha}}^{2\tilde{\alpha}r}}\biggl|
    \begin{array}{c}
     x_1, 1-\mathcal{Z}_1,x_2 \\
     x_3,0 \\
    \end{array}
    \right],
\end{align}
where $\mathcal{Z}_2=\Delta(r,\varepsilon ^2)+ \Delta(r, \alpha_d)+ \Delta(r, \tilde {m}_d)-\Delta(r,\varepsilon ^2+1)-r$, $x_1=\Delta(\tilde{\alpha},1)$, $x_2=\Delta(\tilde{\alpha},l_1)$, and $x_3=\Delta(\tilde{\alpha},l_2)$.
\bibliography{Manuscript}

\begin{thebibliography}{10}
\newcommand{\enquote}[1]{``#1''}

\bibitem{andrews2014will}
J.~G. Andrews, S.~Buzzi, W.~Choi, S.~V. Hanly, A.~Lozano, A.~C. Soong, and
  J.~C. Zhang, \enquote{What will {5G} be?} {\protect\JournalTitle{IEEE Journal
  on selected areas in communications}} \textbf{32}, 1065--1082 (2014).

\bibitem{dang2020should}
S.~Dang, O.~Amin, B.~Shihada, and M.-S. Alouini, \enquote{What should {6G} be?}
  {\protect\JournalTitle{Nature Electronics}} \textbf{3}, 20--29 (2020).

\bibitem{seo2020combined}
S.~Seo, D.-E. Ko, and J.-M. Chung, \enquote{Combined time bound optimization of
  control, communication, and data processing for {FSO}-based {6G UAV} aerial
  networks,} {\protect\JournalTitle{ETRI Journal}} \textbf{42}, 700--711
  (2020).

\bibitem{islam2021impact}
S.~H. Islam, A.~S.~M. Badrudduza, S.~R. Islam, F.~I. Shahid, I.~S. Ansari,
  M.~K. Kundu, and H.~Yu, \enquote{Impact of correlation and pointing error on
  secure outage performance over arbitrary correlated {N}akagami-$m$ and
  {M}-turbulent fading mixed {RF-FSO} channel,} {\protect\JournalTitle{IEEE
  Photonics Journal}}  (2021).

\bibitem{sharma2019effect}
S.~Sharma, A.~Madhukumar, and R.~Swaminathan, \enquote{Effect of pointing
  errors on the performance of hybrid {{FSO/RF}} networks,}
  {\protect\JournalTitle{IEEE Access}} \textbf{7}, 131418--131434 (2019).

\bibitem{1anees2015performance}
S.~Anees and M.~R. Bhatnagar, \enquote{Performance of an amplify-and-forward
  dual-hop asymmetric {RF-FSO} communication system,}
  {\protect\JournalTitle{Journal of Optical Communications and Networking}}
  \textbf{7}, 124--135 (2015).

\bibitem{ansari2013impact}
I.~S. Ansari, F.~Yilmaz, and M.-S. Alouini, \enquote{Impact of pointing errors
  on the performance of mixed {RF/FSO} dual-hop transmission systems,}
  {\protect\JournalTitle{IEEE Wireless Communications Letters}} \textbf{2},
  351--354 (2013).

\bibitem{upadhya2018relay}
A.~Upadhya, V.~K. Dwivedi, and G.~Singh, \enquote{Relay-aided free-space
  optical communications using $\alpha$- $\mu$ distribution over atmospheric
  turbulence channels with misalignment errors,} {\protect\JournalTitle{Optics
  Communications}} \textbf{416}, 117--124 (2018).

\bibitem{7881143}
I.~S. {Ansari}, M.~M. {Abdallah}, M.~{Alouini}, and K.~A. {Qaraqe},
  \enquote{Outage analysis of asymmetric {RF-FSO} systems,} in \emph{2016 IEEE
  84th Vehicular Technology Conference (VTC-Fall),}  (2016), pp. 1--6.

\bibitem{anees2015performance}
S.~Anees and M.~R. Bhatnagar, \enquote{Performance evaluation of
  decode-and-forward dual-hop asymmetric radio frequency-free space optical
  communication system,} {\protect\JournalTitle{IET Optoelectronics}}
  \textbf{9}, 232--240 (2015).

\bibitem{wang2018performance}
Y.~Wang, P.~Wang, X.~Liu, and T.~Cao, \enquote{On the performance of dual-hop
  mixed {RF/FSO} wireless communication system in urban area over aggregated
  exponentiated {W}eibull fading channels with pointing errors,}
  {\protect\JournalTitle{Optics Communications}} \textbf{410}, 609--616 (2018).

\bibitem{6950766}
I.~S. {Ansari}, M.~M. {Abdallah}, M.~{Alouini}, and K.~A. {Qaraqe},
  \enquote{Outage performance analysis of underlay cognitive {RF} and {FSO}
  wireless channels,} in \emph{2014 3rd International Workshop in Optical
  Wireless Communications (IWOW),}  (2014), pp. 6--10.

\bibitem{6952039}
I.~S. {Ansari}, M.~M. {Abdallah}, M.~{Alouini}, and K.~A. {Qaraqe}, \enquote{A
  performance study of two hop transmission in mixed underlay {RF} and {FSO}
  fading channels,} in \emph{2014 IEEE Wireless Communications and Networking
  Conference (WCNC),}  (2014), pp. 388--393.

\bibitem{7883900}
F.~S. {Al-Qahtani}, A.~H.~A. {El-Malek}, I.~S. {Ansari}, R.~M. {Radaydeh}, and
  S.~A. {Zummo}, \enquote{Outage analysis of mixed underlay cognitive {RF MIMO}
  and {FSO} relaying with interference reduction,} {\protect\JournalTitle{IEEE
  Photonics Journal}} \textbf{9}, 1--22 (2017).

\bibitem{erdogan2019performance}
E.~Erdogan, \enquote{On the performance of cognitive underlay {RF/FSO}
  communication systems with limited feedback,} {\protect\JournalTitle{Optics
  Communications}} \textbf{444}, 87--92 (2019).

\bibitem{alimi2017analysis}
I.~A. Alimi, P.~P. Monteiro, and A.~L. Teixeira, \enquote{Analysis of multiuser
  mixed {RF/FSO} relay networks for performance improvements in cloud
  computing-based radio access networks (cc-rans),}
  {\protect\JournalTitle{Optics Communications}} \textbf{402}, 653--661 (2017).

\bibitem{yi2019performance}
X.~Yi, C.~Shen, P.~Yue, Y.~Wang, and Q.~Ao, \enquote{Performance of
  decode-and-forward mixed {RF/FSO} system over $\kappa$- $\mu$ shadowed and
  exponentiated {W}eibull fading,} {\protect\JournalTitle{Optics
  Communications}} \textbf{439}, 103--111 (2019).

\bibitem{yi2020performance}
X.~Yi, Q.~Ao, P.~Yue, C.~Shen, Y.~Wang, and P.~Zhao, \enquote{Performance
  analysis for mixed $\kappa$- $\mu$ shadowed and exponentiated {W}eibull
  distributed dual-hop system with multiuser diversity in c-ran,}
  {\protect\JournalTitle{Optics Communications}} \textbf{460}, 124926 (2020).

\bibitem{zedini2014performance}
E.~Zedini, I.~S. Ansari, and M.-S. Alouini, \enquote{Performance analysis of
  mixed {N}akagami-$ m $ and {G}amma-{G}amma dual-hop {FSO} transmission
  systems,} {\protect\JournalTitle{IEEE Photonics Journal}} \textbf{7}, 1--20
  (2014).

\bibitem{zhao2017performance}
J.~Zhao, S.-H. Zhao, W.-H. Zhao, Y.~Liu, and X.~Li, \enquote{Performance of
  mixed {RF/FSO} systems in exponentiated {W}eibull distributed channels,}
  {\protect\JournalTitle{Optics Communications}} \textbf{405}, 244--252 (2017).

\bibitem{tonk2020mixed}
V.~K. Tonk, A.~Upadhya, P.~K. Yadav, and V.~K. Dwivedi, \enquote{Mixed
  mud-{RF/FSO} two way dcode and forward relaying networks in the presence of
  co-channel interference,} {\protect\JournalTitle{Optics Communications}}
  \textbf{464}, 125415 (2020).

\bibitem{feng2016performance}
J.~Feng and X.~Zhao, \enquote{Performance analysis of mixed {RF/FSO} systems
  with {STBC} users,} {\protect\JournalTitle{Optics Communications}}
  \textbf{381}, 244--252 (2016).

\bibitem{odeyemi2019performance}
K.~O. Odeyemi and P.~A. Owolawi, \enquote{On the performance of transmit
  antenna selection in multiuser asymmetric {RF/FSO} system under generalized
  order user scheduling,} {\protect\JournalTitle{Optik}} \textbf{197}, 163102
  (2019).

\bibitem{wang2019performance}
Z.~Wang, W.~Shi, and W.~Liu, \enquote{Performance analysis of mixed {RF/FSO}
  system with spatial diversity,} {\protect\JournalTitle{Optics
  Communications}} \textbf{443}, 230--237 (2019).

\bibitem{asgari2019performance}
A.~Asgari-Forooshani, M.~Aghabozorgi, E.~Soleimani-Nasab, and M.~A. Khalighi,
  \enquote{Performance analysis of mixed {RF/FSO} cooperative systems with
  wireless power transfer,} {\protect\JournalTitle{Physical Communication}}
  \textbf{33}, 187--198 (2019).

\bibitem{han2018performance}
L.~Han, H.~Jiang, Y.~You, and Z.~Ghassemlooy, \enquote{On the performance of a
  mixed {RF/MIMO FSO} variable gain dual-hop transmission system,}
  {\protect\JournalTitle{Optics Communications}} \textbf{420}, 59--64 (2018).

\bibitem{chen2017multi}
L.~Chen and W.~Wang, \enquote{Multi-diversity combining and selection for
  relay-assisted mixed {RF/FSO} system,} {\protect\JournalTitle{Optics
  Communications}} \textbf{405}, 1--7 (2017).

\bibitem{torabi2019performance}
M.~Torabi and R.~Effatpanahi, \enquote{Performance analysis of hybrid {RF--FSO}
  systems with amplify-and-forward selection relaying,}
  {\protect\JournalTitle{Optics Communications}} \textbf{434}, 80--90 (2019).

\bibitem{zhang2018performance}
Y.~Zhang, X.~Wang, S.-H. Zhao, J.~Zhao, and B.-Y. Deng, \enquote{On the
  performance of 2$\times$ 2 {DF} relay mixed {RF/FSO} airborne system over
  exponentiated {W}eibull fading channel,} {\protect\JournalTitle{Optics
  Communications}} \textbf{425}, 190--195 (2018).

\bibitem{erdogan2019joint}
E.~Erdogan, \enquote{Joint user and relay selection for relay-aided {RF/FSO}
  systems over exponentiated {W}eibull fading channels,}
  {\protect\JournalTitle{Optics Communications}} \textbf{436}, 209--215 (2019).

\bibitem{odeyemi2019selection}
K.~O. Odeyemi and P.~A. Owolawi, \enquote{Selection combining hybrid {FSO/RF}
  systems over generalized induced-fading channels,}
  {\protect\JournalTitle{Optics Communications}} \textbf{433}, 159--167 (2019).

\bibitem{jing2017performance}
Z.~Jing, Z.~Shang-hong, Z.~Wei-hu, and C.~Ke-fan, \enquote{Performance analysis
  for mixed {{FSO/RF}} {N}akagami-$m$ and exponentiated {W}eibull dual-hop
  airborne systems,} {\protect\JournalTitle{Optics Communications}}
  \textbf{392}, 294--299 (2017).

\bibitem{1amirabadi2019performance}
M.~A. Amirabadi and V.~T. Vakili, \enquote{Performance of a relay-assisted
  hybrid {FSO/RF} communication system,} {\protect\JournalTitle{Physical
  Communication}} \textbf{35}, 100729 (2019).

\bibitem{amirabadi2019performance}
M.~A. Amirabadi and V.~T. Vakili, \enquote{On the performance of a multi-user
  multi-hop hybrid {FSO/RF} communication system,}
  {\protect\JournalTitle{Optics Communications}} \textbf{444}, 172--183 (2019).

\bibitem{van2018physical}
B.~Van~Nguyen, H.~Jung, and K.~Kim, \enquote{Physical layer security schemes
  for full-duplex cooperative systems: State of the art and beyond,}
  {\protect\JournalTitle{IEEE Communications Magazine}} \textbf{56}, 131--137
  (2018).

\bibitem{lei2017secrecy}
H.~Lei, Z.~Dai, I.~S. Ansari, K.-H. Park, G.~Pan, and M.-S. Alouini,
  \enquote{On secrecy performance of mixed {RF-FSO} systems,}
  {\protect\JournalTitle{IEEE Photonics Journal}} \textbf{9}, 1--14 (2017).

\bibitem{yang2018physical}
L.~Yang, T.~Liu, J.~Chen, and M.-S. Alouini, \enquote{Physical-layer security
  for mixed $\eta-\mu$ and $\mathcal{M}$-distribution dual-hop {RF/FSO}
  systems,} {\protect\JournalTitle{IEEE Transactions on Vehicular Technology}}
  \textbf{67}, 12427--12431 (2018).

\bibitem{sarker2020secrecy}
N.~A. Sarker, A.~Badrudduza, S.~R. Islam, S.~H. Islam, I.~S. Ansari, M.~K.
  Kundu, M.~F. Samad, M.~B. Hossain, and H.~Yu, \enquote{Secrecy performance
  analysis of mixed hyper-{G}amma and {G}amma-{G}amma cooperative relaying
  system,} {\protect\JournalTitle{IEEE Access}} \textbf{8}, 131273--131285
  (2020).

\bibitem{islam2020secrecy}
S.~H. Islam, A.~Badrudduza, S.~R. Islam, F.~I. Shahid, I.~S. Ansari, M.~K.
  Kundu, S.~K. Ghosh, M.~B. Hossain, A.~S. Hosen, and G.~H. Cho, \enquote{On
  secrecy performance of mixed generalized {G}amma and {M}{\'a}laga {RF-FSO}
  variable gain relaying channel,} {\protect\JournalTitle{IEEE Access}}
  \textbf{8}, 104127--104138 (2020).

\bibitem{lei2020secure}
H.~Lei, H.~Luo, K.-H. Park, I.~S. Ansari, W.~Lei, G.~Pan, and M.-S. Alouini,
  \enquote{On secure mixed {RF-FSO} systems with {TAS} and imperfect {CSI},}
  {\protect\JournalTitle{IEEE Transactions on Communications}} \textbf{68},
  4461--4475 (2020).

\bibitem{lei2018secrecy}
H.~Lei, H.~Luo, K.-H. Park, Z.~Ren, G.~Pan, and M.-S. Alouini, \enquote{Secrecy
  outage analysis of mixed {RF-FSO} systems with channel imperfection,}
  {\protect\JournalTitle{IEEE Photonics Journal}} \textbf{10}, 1--13 (2018).

\bibitem{abd2016security}
A.~H. Abd El-Malek, A.~M. Salhab, S.~A. Zummo, and M.-S. Alouini,
  \enquote{Security-reliability trade-off analysis for multiuser {SIMO} mixed
  {RF/FSO} relay networks with opportunistic user scheduling,}
  {\protect\JournalTitle{IEEE Transactions on Wireless Communications}}
  \textbf{15}, 5904--5918 (2016).

\bibitem{abd2017effect}
A.~H. Abd El-Malek, A.~M. Salhab, S.~A. Zummo, and M.-S. Alouini,
  \enquote{Effect of rf interference on the security-reliability tradeoff
  analysis of multiuser mixed {RF/FSO} relay networks with power allocation,}
  {\protect\JournalTitle{Journal of Lightwave Technology}} \textbf{35},
  1490--1505 (2017).

\bibitem{odeyemi2018physical}
K.~O. Odeyemi and P.~A. Owolawi, \enquote{Physical layer security in mixed
  {RF/FSO} system under multiple eavesdroppers collusion and non-collusion,}
  {\protect\JournalTitle{Optical and Quantum Electronics}} \textbf{50}, 1--19
  (2018).

\bibitem{odeyemi2019security}
K.~O. Odeyemi and P.~A. Owolawi, \enquote{Security outage performance of
  partial relay selection in af mixed {RF/FSO} system with outdated channel
  state information,} {\protect\JournalTitle{Transactions on Emerging
  Telecommunications Technologies}} \textbf{30}, e3555 (2019).

\bibitem{6692669}
I.~S. {Ansari}, F.~{Yilmaz}, and M.~{Alouini}, \enquote{On the sum of squared
  $\eta$-$\mu$ random variates with application to the performance of wireless
  communication systems,} in \emph{2013 IEEE 77th Vehicular Technology
  Conference (VTC Spring),}  (2013), pp. 1--6.

\bibitem{7145973}
I.~S. {Ansari} and M.~{Alouini}, \enquote{On the performance analysis of
  digital communications over {W}eibull-{G}amma channels,} in \emph{2015 IEEE
  81st Vehicular Technology Conference (VTC Spring),}  (2015), pp. 1--7.

\bibitem{10754/134733}
I.~S. Ansari, \enquote{Composite and cascaded generalized-${K}$ fading channel
  modeling and their diversity and performance analysis,}  (2010).

\bibitem{moualeu2018physical}
J.~M. Moualeu, D.~B. da~Costa, W.~Hamouda, U.~S. Dias, and R.~A. de~Souza,
  \enquote{Physical layer security over $\alpha $-$\kappa $-$\mu $ and $\alpha
  $-$\eta $-$\mu $ fading channels,} {\protect\JournalTitle{IEEE Transactions
  on Vehicular Technology}} \textbf{68}, 1025--1029 (2018).

\bibitem{badarneh2015error}
O.~S. Badarneh, \enquote{Error rate analysis of ${M}$-ary phase shift keying in
  $\alpha-\eta-\mu$ fading channels subject to additive {L}aplacian noise,}
  {\protect\JournalTitle{IEEE Communications Letters}} \textbf{19}, 1253--1256
  (2015).

\bibitem{6966082}
I.~S. {Ansari}, M.~{Alouini}, and J.~{Cheng}, \enquote{On the capacity of {FSO}
  links under {L}ognormal and {R}ician-{L}ognormal turbulences,} in \emph{2014
  IEEE 80th Vehicular Technology Conference (VTC2014-Fall),}  (2014), pp. 1--6.

\bibitem{7145711}
I.~S. {Ansari} and M.~{Alouini}, \enquote{Asymptotic ergodic capacity analysis
  of composite {L}ognormal shadowed channels,} in \emph{2015 IEEE 81st
  Vehicular Technology Conference (VTC Spring),}  (2015), pp. 1--5.

\bibitem{salameh2019end}
H.~A.~B. Salameh, L.~Mahdawi, A.~Musa, and F.~H. Tha’er, \enquote{End-to-end
  performance analysis with decode-and-forward relays in multihop wireless
  systems over $\alpha-\eta-\mu$ fading channels,} {\protect\JournalTitle{IEEE
  Systems Journal}} \textbf{14}, 84--92 (2019).

\bibitem{arya2020multiuser}
S.~Arya and Y.-H. Chung, \enquote{Multiuser interference-limited petahertz
  wireless communications over {M}{\'a}laga fading channels,}
  {\protect\JournalTitle{IEEE Access}} \textbf{8}, 137356--137369 (2020).

\bibitem{abramowitz1972handbook}
M.~Abramowitz, I.~A. Stegun \emph{et~al.}, \emph{Handbook of mathematical
  functions: with formulas, graphs, and mathematical tables}, vol.~55 (National
  bureau of standards Washington, DC, 1972).

\bibitem{gradshteyn2014table}
I.~S. Gradshteyn and I.~M. Ryzhik, \emph{Table of integrals, series, and
  products} (Academic press, 2014).

\bibitem{ansari2015performance}
I.~S. Ansari, F.~Yilmaz, and M.-S. Alouini, \enquote{Performance analysis of
  free-space optical links over {M}{\'a}laga ($\mathcal{M}$) turbulence
  channels with pointing errors,} {\protect\JournalTitle{IEEE Transactions on
  Wireless Communications}} \textbf{15}, 91--102 (2015).

\bibitem{saber2018secrecy}
M.~J. Saber and A.~Keshavarz, \enquote{On secrecy performance of mixed
  {N}akagami-$m$ and {M}{\'a}laga {RF/FSO} variable gain relaying system,} in
  \emph{Electrical Engineering (ICEE), Iranian Conference on,}  (IEEE, 2018),
  pp. 354--357.

\bibitem{wyner1975wire}
A.~D. Wyner, \enquote{The wire-tap channel,} {\protect\JournalTitle{Bell system
  technical journal}} \textbf{54}, 1355--1387 (1975).

\bibitem{el2017physical}
A.~H.~A. El-Malek, A.~M. Salhab, S.~A. Zummo, and M.-S. Alouini,
  \enquote{Physical layer security enhancement in multiuser mixed
  \textsc{{RF/FSO}} relay networks under {RF} interference,} in \emph{2017 IEEE
  Wireless Communications and Networking Conference (WCNC),}  (IEEE, 2017), pp.
  1--6.

\bibitem{trinh2016mixed}
P.~V. Trinh, T.~C. Thang, and A.~T. Pham, \enquote{Mixed mmwave {RF/FSO}
  relaying systems over generalized fading channels with pointing errors,}
  {\protect\JournalTitle{IEEE Photonics Journal}} \textbf{9}, 1--14 (2016).

\bibitem{saber2019physical}
M.~J. Saber, A.~Keshavarz, J.~Mazloum, A.~M. Sazdar, and M.~J. Piran,
  \enquote{Physical-layer security analysis of mixed {SIMO SWIPT RF} and {FSO}
  fixed-gain relaying systems,} {\protect\JournalTitle{IEEE Systems Journal}}
  \textbf{13}, 2851--2858 (2019).

\bibitem{lei2018secrecy2}
H.~Lei, Z.~Dai, K.-H. Park, W.~Lei, G.~Pan, and M.-S. Alouini, \enquote{Secrecy
  outage analysis of mixed {RF-FSO} downlink swipt systems,}
  {\protect\JournalTitle{IEEE Transactions on Communications}} \textbf{66},
  6384--6395 (2018).

\bibitem{vellakudiyan2019performance}
J.~Vellakudiyan, V.~Palliyembil, I.~S. Ansari, P.~Muthuchidambaranathan, and
  K.~A. Qaraqe, \enquote{Performance analysis of the decode-and-forward
  relay-based {RF-FSO} communication system in the presence of pointing
  errors,} {\protect\JournalTitle{IET Signal Processing}} \textbf{13}, 480--485
  (2019).

\bibitem{balti2018mixed}
E.~Balti and M.~Guizani, \enquote{Mixed {RF/FSO} cooperative relaying systems
  with co-channel interference,} {\protect\JournalTitle{IEEE Transactions on
  Communications}} \textbf{66}, 4014--4027 (2018).

\bibitem{balti2017aggregate}
E.~Balti, M.~Guizani, B.~Hamdaoui, and B.~Khalfi, \enquote{Aggregate hardware
  impairments over mixed {RF/FSO} relaying systems with outdated {CSI},}
  {\protect\JournalTitle{IEEE Transactions on Communications}} \textbf{66},
  1110--1123 (2017).

\bibitem{palliyembil2018capacity}
V.~Palliyembil, J.~Vellakudiyan, P.~Muthuchidamdaranathan, and T.~A. Tsiftsis,
  \enquote{Capacity and outage probability analysis of asymmetric dual-hop
  {RF--FSO} communication systems,} {\protect\JournalTitle{IET Communications}}
  \textbf{12}, 1979--1983 (2018).

\bibitem{odeyemi2020secrecy}
K.~O. Odeyemi, P.~A. Owolawi, and O.~O. Olakanmi, \enquote{Secrecy performance
  of cognitive underlay hybrid {RF/FSO} system under pointing errors and link
  blockage impairments,} {\protect\JournalTitle{Optical and Quantum
  Electronics}} \textbf{52}, 1--16 (2020).

\bibitem{pattanayak2020physical}
D.~R. Pattanayak, V.~K. Dwivedi, and V.~Karwal, \enquote{Physical layer
  security of a two way relay based mixed {FSO/RF} network in the presence of
  multiple eavesdroppers,} {\protect\JournalTitle{Optics Communications}}
  \textbf{463}, 125429 (2020).

\bibitem{hu2019performance}
J.~Hu, Z.~Zhang, J.~Dang, L.~Wu, and G.~Zhu, \enquote{Performance of
  decode-and-forward relaying in mixed {B}eaulieu-{X}ie and $\mathcal{M}$
  dual-hop transmission systems with digital coherent detection,}
  {\protect\JournalTitle{IEEE Access}} \textbf{7}, 138757--138770 (2019).

\bibitem{badrudduza2020enhancing}
A.~Badrudduza, M.~Sarkar, and M.~Kundu, \enquote{Enhancing security in
  multicasting through correlated {N}akagami-$m$ fading channels with
  opportunistic relaying,} {\protect\JournalTitle{Physical Communication}}
  \textbf{43}, 101177 (2020).

\bibitem{badarneh2015performance}
O.~S. Badarneh and M.~S. Aloqlah, \enquote{Performance analysis of digital
  communication systems over $\alpha$-$\eta$-$\mu$ fading channels,}
  {\protect\JournalTitle{IEEE Transactions on Vehicular Technology}}
  \textbf{65}, 7972--7981 (2015).

\bibitem{gupta2018performance}
J.~Gupta, V.~K. Dwivedi, and V.~Karwal, \enquote{On the performance of {RF-FSO}
  system over {R}ayleigh and $\kappa-\mu$/inverse {G}aussian fading
  environment,} {\protect\JournalTitle{IEEE Access}} \textbf{6}, 4186--4198
  (2018).

\bibitem{badrudduza2021security}
A.~Badrudduza, M.~Ibrahim, S.~R. Islam, M.~S. Hossen, M.~K. Kundu, I.~S.
  Ansari, and H.~Yu, \enquote{Security at the physical layer over {GG} fading
  and m{EGG} turbulence induced {RF-UOWC} mixed system,}
  {\protect\JournalTitle{IEEE Access}} \textbf{9}, 18123--18136 (2021).

\bibitem{prudnikov1988integrals}
A.~P. Prudnikov, Y.~A. Brychkov, O.~I. Marichev, and R.~H. Romer,
  \emph{Integrals and series} (American Association of Physics Teachers, 1988).

\bibitem{mittal1972integral}
P.~Mittal and K.~Gupta, \enquote{An integral involving generalized function of
  two variables,} in \emph{Proceedings of the Indian academy of
  sciences-section A,}  vol.~75 (Springer, 1972), pp. 117--123.

\end{thebibliography}

\end{document}